\documentclass[11pt]{article}
\usepackage{amssymb,amsmath,fullpage,times,epsf}
\usepackage{mathrsfs}
\usepackage{amssymb}
\usepackage{graphics}
\usepackage{graphicx}
\usepackage{subfigure}
\usepackage{float}
\usepackage{bm}
\usepackage{amsfonts}
\usepackage{color}

\allowdisplaybreaks

\newcommand{\mi}{\mathrm{i}}

\def\lam{\lambda}
\def\mb{\mathbf}
\def\ra{\rightarrow}

\def\lim{\mathrm{lim}}

\def\sgn{\mathrm{sgn}}

\def\PT{$\mathcal{P}\mathcal{T}$}

\def\eref#1{(\ref{#1})}

\newcommand{\tabincell}[2]{\begin{tabular}{@{}#1@{}}#2\end{tabular}}

\usepackage[square,numbers,sort&compress]{natbib}
\begin{document}

\date{}

\title{Mixed soliton solutions of the defocusing nonlocal nonlinear Schr\"{o}dinger equation}

\author{Tao Xu$^{1,2,}$\thanks{Corresponding author, e-mail: xutao@cup.edu.cn}\,, Sha Lan$^{2}$,
Min Li$^{3,}$\thanks{Corresponding author, e-mail: micheller85@126.com}\,, Ling-Ling Li$^{2}$, Guo-Wei Zhang$^{2}$
\\
{\em 1. State Key Laboratory of Heavy Oil Processing,}\\
{\em China University of Petroleum, Beijing 102249, China }\\
{\em 2. College of Science, China University of Petroleum, Beijing
102249, China}
\\{\em 3.  School of Mathematics and Physics,}\\
{\em  North China Electric Power University, Beijing 102206,
China}}
\maketitle
\vspace{-5mm}

\begin{abstract}

By using the Darboux transformation, we obtain two new types of exponential-and-rational mixed soliton solutions for the defocusing nonlocal nonlinear Schr\"{o}dinger equation. We reveal that the first type of solution can display a large variety of interactions among two exponential solitons and two rational solitons, in which the standard elastic interaction properties are preserved and each soliton could be either the dark or antidark type. By developing the asymptotic analysis method, we also find that the second type of solution can exhibit the elastic interactions among four mixed asymptotic solitons. But in sharp contrast to the common solitons, the  mixed asymptotic solitons have the $t$-dependent velocities and their phase shifts  before and after interaction also grow with $|t|$  in the logarithmical manner. In addition, we discuss the degenerate cases for such two types of mixed soliton solutions when the four-soliton interaction reduces to a three-soliton  or two-soliton interaction.

\vspace{5mm}


\end{abstract}

\newpage

\section{Introduction}

Recently, it  has become an active topic to study the integrable nonlocal evolution equations in nonlinear mathematical physics and soliton theory. In 2013, Ablowitz and Musslimani first proposed the following  nonlocal nonlinear Schr\"{o}dinger (NLS) equation~\cite{Ablowitz1}:
\begin{align}
 i u_t = u_{xx} + 2 \varepsilon u^2\bar{u}
\quad (\varepsilon= \pm 1)\,, \label{NNLS}
\end{align}
where $u$ is a complex-valued function of real variables $x$ and $t$, $\varepsilon=1$ and $\varepsilon=-1$ represent, respectively,
the focusing $(+)$ and defocusing $(-)$ nonlinearity, and the $bar$ denotes the combination of complex conjugate and space reversal,
 i.e., $\bar{u}=u^*(-x,t)$.  
Compared with the standard (local) NLS equation, the cubic nonlinear term  $|u|^2u $ is replaced with  $u^2\bar{u}$,  so that  the evolution dynamics of the field in Eq.~\eref{NNLS} is  non-locally dependent on the values of $u$  at both the positions $x$ and $-x$.  Also, this equation is said to be
parity-time ($\mathcal{P}\mathcal{T}$) symmetric since it is invariant under the combined action of parity operator $\mathcal{P}$ ($x \ra -x$) and
time-reversal  operator $\mathcal{T}$ ($t \ra -t, i \ra -i$). As a matter of fact,  Eq.~\eref{NNLS} can be viewed as a linear \PT-symmetric
Schr\"{o}dinger equation $i\,u_t = u_{xx} + V(x,t)u $ with the self-induced  potential $V(x,t)=2\, \varepsilon\,u \bar{u}$ satisfying the \PT-symmetric
relation $V(x,t)=V^{*}(-x,t)$. This makes Eq.~\eref{NNLS} relate to the  $\mathcal{P}\mathcal{T}$-symmetric physics~\cite{PTPhys}, and may bring potential applications in some unconventional physical systems~\cite{Sarma,Magstr}.

It is remarkable that Eq.~\eref{NNLS} is a Hamiltonian integrable model in the sense of admitting the Lax pair and an infinite
number of conservation laws, and that its initial-value problem can be solved  via the inverse scattering transform (IST)~\cite{Ablowitz1,Ablowitz3,Ablowitz4,Ablowitz5,Ablowitz6,Ablowitz7}. In the past few years, this equation has  attracted intensive attention and its integrable properties and solution dynamics have been studied from different points of view. Some progresses have been made in the following aspects: the IST scheme for solving Eq.~\eref{NNLS} with nonzero boundary conditions~\cite{Ablowitz5}, gauge equivalence to the unconventional system of coupled Landau-Lifshitz
equations~\cite{Magstr}, complete integrability of a whole hierarchy of nonlocal NLS equations~\cite{Gerdjikov},
long-time asymptotic behavior of the solution with  decaying  boundary conditions~\cite{Rybalkoy},
connection between nonlocal and local NLS equations through the variable transformation~\cite{YJK1},
exact soliton and rogue-wave solutions by different analytical methods~\cite{Ablowitz1,Ablowitz3,Ablowitz4,Ablowitz5,Sarma,LiXu,LiXu1,LiXu2,HJS,Gupta,YJK2,YJK3,YZY,Wen,Gurses,Khare,ZDJ,FBF,LML}.
In addition,  by considering the parity/time reversal, charge conjugation, space/time translation  and their proper combinations,   many other integrable nonlocal models have been obtained from their local counterparts. Typical examples include the
nonlocal reverse space-time NLS equation~\cite{Ablowitz7}, semi-discrete  NLS equation~\cite{Ablowitz2}, vector (or multi-component) NLS equation~\cite{vector,Sinha}, derivative NLS equation~\cite{DNLS,NKN}, modified Korteweg-de Vries equation~\cite{Ablowitz3,Ablowitz4,mKdV},
sine-Gordon equation~\cite{Ablowitz3,Ablowitz4,Ablowitz6}, Davey-Stewartson (DS) equation~\cite{Fokas,Ablowitz4},
$N$-wave system~\cite{NNWave}, Sasa-Satsuma equation~\cite{SS} and various nonlocal Alice-Bob systems~\cite{AB-KdV,Lou2}.

As shown in the previous studies~\cite{Ablowitz1,Ablowitz3,Ablowitz4,Ablowitz5,Sarma,LiXu,LiXu1,LiXu2,HJS,Gupta,YJK2,YJK3,YZY,Wen,Gurses,Khare,FBF,LML}, the nonlocal NLS equation possesses abundant localized-wave solutions and their dynamical behaviors are distinguished from those in the local counterpart. The focusing nonlocal NLS equation has  the bright-soliton, dark-soliton, rogue-wave and breather solutions~\cite{Ablowitz1,Sarma,Khare,Gupta,YJK2,YJK3}, simultaneously. Those solutions are bounded only for some particular parametric choices, but they in general develop the collapsing singularities in
finite time~\cite{Ablowitz1,Ablowitz3,YJK2,YJK3}. For the defocusing case, Eq.~\eref{NNLS} admits the exponential soliton solutions as well as the rational soliton solutions on the same continuous wave (cw) background $u_{\rm{cw}}=\rho\,e^{i\,(2 \rho^2 t +\phi)}$ (with $\rho\neq 0$ and $\phi $ being two real parameters)~\cite{LiXu,LiXu1,LiXu2,Wen,HJS,Ablowitz5}. Both types of soliton solutions can display a rich variety of elastic interactions and each asymptotic soliton could be either the dark or antidark type. In contrast to the standard elastic soliton interactions, some unusual interaction properties have been revealed: the exponential $N$-soliton solution contains generally $2N$ interacting solitons~\cite{LiXu}; the rational soliton experiences no phase shift when interacting with another rational one~\cite{LiXu1}; the asymptotic solitons in the higher-order rational solutions  have the $t$-dependent velocities because their center trajectories are localized in some curves in the $xt$ plane~\cite{LiXu2}. However, whether for  the focusing or defocusing nonlocal NLS equation, the  stability of localized-wave solutions can  be easily destroyed if there is a small shift for the initial value in the $x$ coordinate~\cite{Sarma,LiXu,LiXu1}.

It is known that the Darboux transformation (DT) is an algebraic iterative method which can generate an infinite chain of explicit solutions for the Lax integrable equations from a trivial seed~\cite{Matveev1}. Lately, we  have succeeded in  constructing the exponential and rational soliton solutions of the defocusing nonlocal NLS equation~\cite{LiXu,LiXu1} by the elementary and generalized DTs. If the potential is given by the cw solution $\rho\,e^{i\,(2 \rho^2 t +\phi)}$, the Lax pair of Eq.~\eref{NNLS} with $\varepsilon=-1$ usually has the solution in the exponential form (see Eq.~\eref{lpsola} below), but such solution will reduce to a rational one (see Eq.~\eref{lpsolb} below) when the spectral parameter takes the critical value $\lam= i \sigma \rho $ ($\sigma=\pm 1$). As a result, the elementary DT can be used to derive the exponential soliton solutions based on a set of $N$ linearly independent solutions at different spectral parameters $\{\lam_k\}_{k=1}^N$ with $\lam_k\neq i \sigma \rho$~\cite{LiXu}, while the generalized DT can generate the rational soliton solutions when all $\lam_k$'s degenerate to $i \sigma \rho$~\cite{LiXu1}. However, it is also possible for $N$ spectral parameters to partially degenerate to $i \sigma \rho$ or the degeneration occurs at any non-critical value. That is to say, quite a number of degenerate cases have been overlooked in the existing literature although they can still be dealt with via the generalized DT.  With this consideration at $N=2$, we in this paper construct two new types of exponential-and-rational mixed soliton solutions for the defocusing nonlocal NLS equation. We reveal that the first type of solution can display a large variety of elastic interactions, in which there are in general two exponential solitons and two rational solitons, and each interacting soliton could be either the dark or antidark type. For the second type of solution, we develop the asymptotic analysis method and find that the solutions contain four mixed  asymptotic solitons and each one can also display the  dark or antidark soliton profile. Very specially, all the center trajectories of mixed solitons are localized in some curves in the $xt$ plane, so that they have the $t$-dependent velocities and their phase shifts before and after interaction grow with $|t|$ in a logarithmical manner.

The structure of this paper is organized as follows: In Section~\ref{Sec2}, we  review the elementary and generalized DTs of Eq.~\eref{NNLS}, as proposed in Refs.~\cite{LiXu,LiXu1}. In Section~\ref{Sec3}, for the case $\lam_1 = i b \rho$  ($0 <|b| < 1 $) and  $\lam_2 = i \sigma \rho$, we use the elementary  DT to construct the first type of mixed soliton solution, and reveal the elastic soliton interaction properties through an asymptotic analysis. In Section~\ref{Sec4}, for another case $\lam_2, \lam_1 \ra  i b \rho$  ($0 <|b| < 1 $), we derive the second type of mixed soliton solution by the generalized DT. Specially, we develop the asymptotic analysis method and obtain some uncommon soliton interaction properties which have never been reported before. In Section~\ref{Sec5}, we address the conclusions and discussions of our work.

\section{Darboux Transformation}
\label{Sec2}

As a special gauge transformation leaving the form of Lax pair invariant, the DT comprises of the eigenfunction and potential transformations~\cite{Matveev1,GuHu}. Since Eq.~\eref{NNLS} is an integrable model,  it has the Lax pair in the form~\cite{Ablowitz1}:
\begin{subequations}
\begin{align}
& \Psi_x=U\Psi=\begin{pmatrix}
\lam & u(x,t) \\
-\varepsilon u^*(-x,t) & -\lam
\end{pmatrix}\Psi,\label{CNLS4a} \\
& \Psi_t=V\Psi=\begin{pmatrix}
-2 i \lam^2- i \varepsilon u(x,t) u^*(-x,t) & -2\, i \lam u(x,t) - i u_x(x,t)\\
2  i \varepsilon \lam   u^*(-x,t) -  i \varepsilon u_x^*(-x,t) &
2 i \lam^2 +  i \varepsilon u(x,t) u^*(-x,t)
\end{pmatrix}\Psi,\label{CNLS4b}
\end{align}
\end{subequations}
where $ \Psi=(f, g)^{\rm{T}}$ (the superscript $\rm{T}$
represents the vector transpose) is the vector eigenfunction, $
\lambda $ is the spectral parameter, and Eq.~\eref{NNLS} can be
recovered from the compatibility condition $U_t - V_x + U\,V -
V\,U=0$.

Assume that $\Psi_k=\big[f_k(x,t), g_k(x,t)]^{\rm{T}}$ ($1\leq k\leq N$)
are $N$ linearly-independent solutions of Eqs.~\eref{CNLS4a} and~\eref{CNLS4b} with different spectral parameters $\lam_k$ ($ 1\leq k\leq N$),
where $\lam_k$ cannot be taken as  a real number  to avoid the trivial  iteration  of the DT.
One  can check that $\bar{\Psi}_k=\big[g^*_k(-x,t), \varepsilon
f^*_k(-x,t)\big]^{\rm{T}}$ also solves Eqs.~\eref{CNLS4a} and~\eref{CNLS4b} with $\lam = \lam^*_k$.
Based on the work in Ref.~\cite{LiXu}, the  $N$th-iterated elementary DT  can be
constituted by the eigenfunction transformation
\begin{align}
& \Psi_{[N]}=T_{[N]}\Psi, \quad   T_{[N]}=
\begin{pmatrix}
\lam^N  - \sum\limits_{n=1}^{N}a_n(x,t) \lam^{n-1}  & - \sum\limits_{n=1}^{N}b_n(x,t) (-\lam)^{n-1}  \\
- \sum\limits_{n=1}^{N}c_n(x,t) \lam^{n-1}  & \lam^N  -
\sum\limits_{n=1}^{N}d_n(x,t)
(-\lam)^{n-1}   \\
\end{pmatrix}  \label{ETR}
\end{align}
and the potential transformation
\begin{align}
&  u_{[N]}(x,t) = u(x,t) +2\,(-1)^{N-1} b_{N}, \quad u^*_{[N]}(-x,t)=
u^*(-x,t) + 2\,\varepsilon c_{N}. \label{PotentialTrana}
\end{align}
The functions $ a_n(x,t), b_n(x,t), c_n(x,t)$ and $d_n(x,t)$ $( 1\leq  n \leq N )$ are uniquely determined by
\begin{align}
& T_{[N]}\mid_{\lam=\lam_k}\!\Psi_k=\mb{0}, \quad
T_{[N]}\mid_{\lam=\lam^*_k}\!\bar\Psi_k=\mb{0} \quad (1 \leq k \leq N),
\label{UndetCoeff}
\end{align}
and particularly $b_N$ and $c_N$ can be represented  as
\begin{align}
& b_N= (-1)^{N-1}\frac{\tau_{N+1,N-1}}{\tau_{N,N}},\quad c_N =
 \frac{\tau_{N-1,N+1}}{\tau_{N,N}},
\label{PotentialTranb}
\end{align}
with
\begin{align}
 \tau_{M, L}=
\begin{vmatrix}
F_{N\times M} & G_{N\times L} \\
\varepsilon \bar{G}_{N\times M} & \bar{F}_{N\times L}
\end{vmatrix} \quad (M+L=2N),
\end{align}
where the block matrices $ F_{N\times M} =
\big[\lam_k^{m-1}f_k(x,t)\big]_{1\leqslant k \leqslant N, \atop 1\leqslant m \leqslant M}$,
$G_{N\times L}= \big[(-\lam_k)^{m-1}g_k(x,t)\big]_{1\leqslant k \leqslant N, \atop 1\leqslant m \leqslant L}$,
$\bar{G}_{N\times M} =
\big[\lam^{*m-1}_k g^*_k(-x,t)\big]_{1\leqslant k \leqslant N, \atop 1\leqslant m \leqslant M}$ and $\bar{F}_{N\times L} =
\big[(-\lam^*_k)^{m-1}f^*_k(-x,t)\big]_{1\leqslant k \leqslant N, \atop 1\leqslant m \leqslant L}$.

We notice that the elementary DT cannot apply to the degenerate cases when some of the spectral parameters $\{\lam_k\}_{k=1}^N$ coincide with each other because the coefficient matrix in Eq.~\eref{UndetCoeff} becomes singular. This difficulty may be overcome by the  idea of Matveev's generalized DT~\cite{Matveev,GDT}. Let us consider the following general case:
\begin{align}
\lam_{k_1+1}, \dots, \lam_{k_2-1}\ra \lam_{k_1}, \,
\lam_{k_2+1}, \dots,\lam_{k_3-1} \ra \lam_{k_2},  \cdots,
\lam_{k_n+1}, \dots, \lam_N \ra \lam_{k_n}, \label{Degcase}
\end{align}
where $1=k_1 < k_2 < \dots < k_n \leq N $ ($\lam_{k_i} \neq \lam_{k_j} $, $1\leq i <j \leq n$, $1\leq n \leq N$).
For convenience, we define that $l_i= k_{i+1}-k_i-1$ for $1\leq i \leq n-1 $ and
$l_n=N-k_n$, and assume that
\begin{align}
f_{k_i+h} (x,t)=
f_{k_i}(x,t,\lam_{k_i+h} ), \quad
g_{k_i+h}(x,t) =
g_{k_i}(x,t,\lam_{k_i+h} ),
\end{align}
where $\lam_{k_i+h} = \lam_{k_i} + \epsilon_i $,  $ 1\leq h \leq l_i$ ($l_i>0$),
$\epsilon_i$'s are small parameters. By expanding $\lam_{k_i+h}^{m-1} f_{k_i} $ and $\lam_{k_i+h}^{m-1} g_{k_i}$ in the
Taylor series of $\epsilon_i$ and taking the limit $\epsilon_i \ra 0$, the functions $ a_n(x,t)$, $b_n(x,t)$, $c_n(x,t)$  and $d_n(x,t)$  in $T_{[N]}$ can be uniquely solved from  Eq.~\eref{UndetCoeff} again.
As a result, the  potential transformation is replaced by
\begin{align}
u_{[N]}(x,t) =u(x,t)+2\,\frac{\tau'_{N+1,N-1}}{\tau'_{N,N}}, \quad
u^*_{[N]}(-x,t)= u^*(-x,t) +2\,\varepsilon\frac{\tau'_{N-1,N+1}}{\tau'_{N,N}},\label{NPT}
\end{align}
with \begin{align}
\tau'_{M, L}=
\begin{vmatrix}
F'_{l_1\times M} & G'_{l_1\times L} \\
\vdots  &   \vdots  \\
F'_{l_n\times M}  &  G'_{l_n\times L}  \\
\varepsilon \bar{G}'_{l_1\times M} & \bar{F}'_{l_1\times L} \\
\vdots  &   \vdots  \\
\varepsilon \bar{G}'_{l_n\times M} & \bar{F}'_{l_n\times L}
\end{vmatrix}\quad (M+L=2N),\label{PTR}
\end{align}
in which the block matrices $ F'_{l_i\times M}=\left[f_{k_i}^{(m-1,j)}(x,t)\right]_{0 \leq j \leq l_i,  \atop 1 \leq m \leq M}$, $G'_{l_i\times L}=\left[ (-1)^{m-1} g_{k_i}^{(m-1,j)}(x,t)\right]_{0 \leq j \leq l_i, \atop 1 \leq m \leq L}$, $\bar{G}'_{l_i\times M}=\left[ g_{k_i}^{*(m-1,j)}(-x,t)\right]_{0 \leq j \leq l_i,  \atop 1 \leq m \leq M}$ and $ \bar{F}'_{l_i\times L}=\left[(-1)^{m-1}
f_{k_i}^{*(m-1,j)}(-x,t)\right]_{0 \leq j \leq l_i, \atop 1 \leq m \leq L}$ ($1\leq i \leq n$), and the functions $f_{k_i}^{(m-1,j)} $ and $g_{k_i}^{(m-1,j)}$ are defined by
\begin{align}
& f_{k_i}^{(m-1,j)}(x,t) =\frac{1}{j !}\frac{\partial^j
[\lam_{k_i+h}^{m-1}\, f_{k_i} (x,t, \lam_{k_i+h})]}{\partial
\lam^j_{k_i+h}}\Big|_{\epsilon_i=0}, \label{Taylor1} \\
& g_{k_i}^{(m-1,j)}(x,t) =\frac{1}{j !}\frac{\partial^j
[\lam^{m-1}_{k_i+h}\, g_{k_i} (x,t, \lam_{k_i+h})]}{\partial
\lam_{k_i+h}^j}\Big|_{\epsilon_i=0}, \label{Taylor2}
\end{align}
where $1\leq i \leq n $, $1\leq m\leq N$, $1\leq h \leq l_i$, and
particulary $f_{k_i}^{(0,0)}(x,t)=f_{k_i}(x,t)$,
$g_{k_i}^{(0,0)}(x,t)=g_{k_i}(x,t)$.

Therefore, we call Eqs.~\eref{ETR} and~\eref{NPT}  the  $N$th-iterated generalized DT, which is applicable to any choice of the spectral parameters $\{\lam_k\}_{k=1}^N$ only if $\lam_k \neq \lam^*_k$. One should note that the  potential transformation in Eq.~\eref{PotentialTrana} corresponds to the particular case of the generalized one in Eq.~\eref{NPT} when $n=N$.

With the cw solution $u_{\rm{cw}}=\rho\,e^{i\,(2 \rho^2 t +\phi)}$ (where $\rho\neq 0$ and $\phi $ are two real parameters) as a seed, we implement the DT-iterated algorithm for Eq.~\eref{NNLS} with $\varepsilon=-1$. In this case, depending on the value of the spectral parameter,  the Lax pair~\eref{CNLS4a} and~\eref{CNLS4b} has two different solutions:
\begin{align}
&
\begin{pmatrix}
f_k\\
g_k
\end{pmatrix}=
\begin{pmatrix}
 e^{\frac{ 2\, i  \rho^2 t + i \phi}{2}}\big(\alpha_k  e^{ h_k  \xi_k}+\beta_k  e^{ -h_k  \xi_k }\big)\\
 e^{-\frac{ 2\, i  \rho^2 t + i \phi}{2}}\big[\frac{\alpha_k(h_k-
\lam_k)}{\rho}  e^{h_k \xi_k } - \frac{\beta_k(h_k + \lam_k
)}{\rho}  e^{ -h_k \xi_k }\big]
\end{pmatrix} \quad (\lam_k \neq   i \sigma \rho), \label{lpsola} \\
& \begin{pmatrix} f_k\\ g_k
\end{pmatrix}=
\begin{pmatrix}
 e^{\frac{ 2\, i  \rho^2 t + i \phi}{2}}\big[\alpha _k (x+ 2\,\sigma\rho t)+\beta _k \big]\\
 e^{-\frac{ 2\, i  \rho^2 t + i \phi}{2}}\big[- i \alpha _k \sigma  (x+2 \sigma \rho  t  ) + \frac{\alpha _k}{\rho }- i \beta _k \sigma \big]
\end{pmatrix} \quad\,\,\,\,\, (\lam_k =  i \sigma \rho), \label{lpsolb}
\end{align}
where $\sigma = \pm1$, $h_k =\sqrt{\lam_k^2 + \rho^2}$, $\xi_k =x- 2\, i  \lam_k
t$,  and  $ \alpha_k$ and $\beta_k $  are free
complex parameters.  If taking $\lam_k =  i b_k \rho $ with $0 <|b_k| < 1 $ and $b_k \neq b_j $ for all $1\leq k \leq  N$, the potential transformation~\eref{PotentialTrana} can give rise to a chain of  exponential soliton solutions~\cite{LiXu}.  Instead, if $ \lam_2, \dots, \lam_N \ra \lam_1=  i \sigma \rho $ (which corresponds to $n=1$ and $l_1=N-1$ in Eq.~\eref{Degcase}),  one can derive the  rational soliton solutions from Eq.~\eref{NPT}~\cite{LiXu1}. It should be noted that Eq.~\eref{Degcase} contains quite a number of other degenerate cases which have been overlooked in the previous studies. With $N=2$ as an example, there are  the following two cases remaining to be studied:  (i) $\lam_1 = i b \rho$ ($0 <|b| < 1 $), $\lam_2 =  i \sigma \rho$; (ii) $\lam_2, \lam_1 \ra  i b \rho$  ($0 <|b| < 1 $). In Sections~\ref{Sec3} and~\ref{Sec4}, by considering such two degenerate cases, we will derive two new types of mixed soliton solutions and discuss the soliton interaction properties via asymptotic analysis.

\section{The first type of mixed soliton solution}
\label{Sec3}

In this section, by letting $ \lam_1 = i \, b \rho $ ($0 <|b| < 1 $) and $ \lam_2 =  i \, \sigma  \rho$  ($\sigma=\pm1$), we use the  elementary DT to obtain the exponential-rational mixed soliton solution as follows:
\begin{align}
u= \rho  e^{2\, i \rho^2 t + i \phi} + 2 \frac{\begin{vmatrix}
f_1 & \lambda_1 f_1 & \lambda_1^2 f_1 &  g_1  \\
f_2 & \lambda_2 f_2 & \lambda_2^2 f_2 &  g_2 \\
-\bar{g}_1 &  - \lambda^*_1 \bar{g}_1 &  - \lambda^{*2}_1 \bar{g}_1 & \bar{f}_1 \\
-\bar{g}_2 &  - \lambda^*_2 \bar{g}_2 &  - \lambda^{*2}_2 \bar{g}_2 & \bar{f}_2
\end{vmatrix}}{
\begin{vmatrix}
f_1 & \lambda_1 f_1 &  g_1 & -\lambda_1 g_1  \\
f_2 & \lambda_2 f_2 &  g_2 & -\lambda_2 g_2 \\
- \bar{g}_1 &  - \lambda^*_1 \bar{g}_1  & \bar{f}_1 &  - \lambda^*_1 \bar{f}_1   \\
- \bar{g}_2 &  - \lambda^*_2 \bar{g}_2  & \bar{f}_2 &  - \lambda^*_2 \bar{f}_2
\end{vmatrix}}, \label{2solution}
\end{align}
with
\begin{subequations}
\begin{align}
& f_1=  \alpha_1 e^{ i \rho^2 t +\frac{ i \phi}{2}}( e^{\kappa \xi_1} + \gamma_1 e^{-\kappa \xi_1}),\\
&g_1=  \alpha_1 e^{- i  \rho^2 t -\frac{ i  \phi }{2}}(  e^{\kappa \xi_1- i \theta}  - \gamma_1 e^{-\kappa \xi_1+ i \theta}), \\
&f_2=\alpha_2  e^{ i  \rho^2 t +\frac{ i  \phi }{2}}(\eta_1+\gamma_2), \\
&g_2 = -\frac{i \alpha_2}{\rho}  \, e^{- i \rho^2 t -\frac{ i  \phi }{2}} [ \sigma \rho  (\eta_1+\gamma_2) + i],  \\
&  \xi_1=x+2b \rho  t, \quad \eta_1=x+2\sigma \rho  t, \quad \kappa=\rho\sqrt{1-b^2}, \\
& \theta= \arctan(b/\sqrt{1-b^2}), \quad  \gamma_k=\beta_k/\alpha_k \quad(k=1,2),
\end{align}
\end{subequations}
where $\bar{f}_k =f^*_k(-x,t),\,\bar{g}_k =g^*_k(-x,t)$ ($k=1,2$),  $-\pi/2 < \theta< \pi/2$, $\gamma_1$ and $\gamma_2$ are two complex parameters. Note that $\alpha_1$ and $\alpha_2$ will be canceled out when $f_k$ and $g_k$ ($k=1,2$) are substituted into  Eq.~\eref{2solution}. For convenience, we define $\gamma_1= r_1 e^{ i \varphi_1}$ with  $-\pi< \varphi_1 \leq \pi$ and $r_1>0$ being a real constant.

\subsection{Asymptotic analysis}

We use the asymptotic analysis method to study the soliton interactions described by  solution~\eref{2solution}. It turns out that solution~\eref{2solution} has four different asymptotic soliton states when $|t|\ra \infty$, which are given as follows:
\begin{enumerate}
\item[(i)]
If $\xi_1=x+2b \rho  t= O(1)$,  from $\bar\xi_1=-\xi_1 + 4b \rho  t $ we have $\bar\xi_1 \ra  \pm \infty $ as $\sgn(b) t \ra  \pm \infty$.
Then,  calculating the limit of solution~\eref{2solution} when $\bar\xi_1\ra\pm\infty$ gives  the following asymptotic expression in the exponential form
\begin{subequations}
\begin{align}
& u\ra \left\{
\begin{array}l
u_{1}^{+}=\rho  e^{2\, i \rho^2 t + i \phi}\Big[1 + \dfrac{i(1 + e^{2 i \theta})}{b\,r_1 e^{-2\kappa \xi_1 + i (\theta+\varphi_1) } - i }\Big] \quad(\sgn(b) t \ra \infty), \\[1mm]
u_{1}^{-}=\rho  e^{2\, i \rho^2 t + i \phi}\Big[1 - \dfrac{i(1 + e^{-2 i \theta})}{b\,r^{-1}_1 e^{2\kappa \xi_1 - i (\theta+\varphi_1) } + i }\Big] \quad(\sgn(b) t \ra -\infty),
\end{array} \right.  \label{1asymptotica} \\[2mm]
&  |u|\ra |u_{1}^{\pm}|=\rho\Big[1 + \frac{2\sqrt{1-b^2} \sin(\varphi_1) }{
\sgn(b)\cosh(2\kappa \xi_1+\Delta_1^{\pm})- \sin \left(\theta+\varphi _1\right) } \Big]^{\frac{1}{2}}, \label{1asymptoticb}
\end{align}
\end{subequations}
with $\Delta_1^{\pm}=-\ln\big(r_1|b|^{\pm\!1}\big)$.
It can be seen from Eq.~\eref{1asymptoticb} that $u_{1}^{\pm}$ has no singularity
if and only if $\gamma_1$ and $b$ satisfy
\begin{align}
&  \sin \left(\theta+\varphi _1\right) \neq \sgn(b), \label{cond1}
\end{align}
and it is localized in the line $ 2\kappa \xi_1+\Delta_1^{\pm}=0$.  For $b \sin \left(\varphi _1\right) >0$ and $b \sin \left(\varphi _1\right)<0$, $u_{1}^{\pm}$ can, respectively,  represent an exponential antidark (EAD)  soliton on top of the cw background $u=\rho\, e^{2\, i \rho^2 t + i \phi}$ and an exponential dark (ED) soliton beneath the same background.

\item[(ii)]
If $\bar\xi_1= -x+2b \rho  t= O(1)$,  from $\xi_1=-\bar\xi_1 + 4b \rho  t $ we also have $\xi_1 \ra  \pm \infty $ as $\sgn(b) t \ra  \pm \infty$.
Then,  calculating the limit of solution~\eref{2solution} when $\xi_1\ra\pm\infty$ gives another asymptotic expression in the exponential form
\begin{subequations}
\begin{align}
& u \ra \left\{
\begin{array}l
u_{2}^+=\rho  e^{2\, i \rho^2 t +  i \phi} \Big[1  -
\dfrac{i(1 + e^{2 i \theta})}{b\,r_1 e^{-2\kappa \bar{\xi}_1 - i (\theta+\varphi_1) } +  i }\Big] \quad(\sgn(b) t \ra \infty),  \\[1mm]
u_{2}^-= \rho  e^{2\, i \rho^2 t +  i \phi} \Big[1 +
\dfrac{i(1 + e^{-2 i \theta})}{b\,r^{-1}_1 e^{2\kappa \bar{\xi}_1 + i (\theta+\varphi_1) } -  i }\Big]\quad(\sgn(b) t \ra -\infty),
\end{array} \right. \label{2asymptotica} \\[2mm]
& |u| \ra |u_{2}^{\pm}|=\rho\Big[1 + \frac{2\sqrt{1-b^2}\,\sin \left(2\theta +\varphi _1\right)}{\sgn(b) \cosh(- 2 \kappa \bar{\xi}_1 - \Delta_1^{\pm})- \sin \left(\theta +\varphi _1\right) }\Big]^{\frac{1}{2}}.  \label{2asymptoticb}
\end{align}
\end{subequations}
When condition~\eref{cond1} is satisfied, $u_{2}^{\pm}$ is also nonsingular and is located in the line $ 2\kappa \bar{\xi}_1 + \Delta_1^{\pm} =0$. Likewise, $u_{2}^{\pm}$ can  display the EAD and ED soliton profiles  which are  associated with $b \sin \left(2\theta +\varphi _1\right) >0$ and $b \sin \left(2\theta +\varphi _1\right)<0$, respectively.


\item[(iii)]
If $\eta_1= x+2\sigma \rho  t= O(1)$, from $\xi_1 = \eta_1 + 2(b-\sigma)\rho t $ and $\bar\xi_1 =-\eta_1 + 2(b+\sigma)\rho t $  we know that $\xi_1 \ra \mp \infty$ and $\bar\xi_1 \ra \pm \infty$ as $\sigma t\ra \pm \infty$. Then, by taking the limit of of solution~\eref{2solution} when $\xi_1,-\bar\xi_1\ra -\infty$ and $\xi_1,-\bar\xi_1\ra \infty$, we have the following asymptotic expression in the rational form
\begin{subequations}
\begin{align}
&u\ra  u_{3}^{\pm}=\rho  e^{2\, i \rho^2 t + i \phi}\bigg[1+\frac{2\,\sigma }{2\, i \rho  (\eta_1+\gamma_2) -\sigma \pm 2\, i \sec(\theta) }\bigg] \quad (\sigma t\ra \pm \infty), \label{3asymptotica}\\
&|u| \ra  |u_{3}^{\pm}|=\rho\bigg\{1- \frac{8\,\rho\, \sigma \gamma_{2I} }{4 \big[\rho \eta_1+\rho \gamma_{2R}\pm \sec(\theta) \big]^2+(2\rho \gamma_{2I}+\sigma)^2}\bigg\}^{\frac{1}{2}}. \label{3asymptoticb}
\end{align}
\end{subequations}
Apparently, $u_{3}^{\pm}$  has no singularity if and only if
\begin{align}
 2\rho \gamma_{2I}+\sigma \neq 0, \label{cond2}
\end{align}
and it can describe an rational antidark (RAD) soliton for $\sigma  \gamma_{2I}<0$ or an rational dark (RD) soliton  for $\sigma \gamma_{2I}>0$ in the line $ \eta_1+ \gamma_{2R} \pm \frac{\sec(\theta)}{\rho } =0$.

\item[(iv)]
If $\bar\eta_1= -x+2\sigma \rho  t= O(1)$,  from $\xi_1 = -\bar\eta_1 + 2(b+\sigma)\rho t $ and $\bar\xi_1 =\bar\eta_1 + 2(b-\sigma)\rho t $ we have
$\xi_1 \ra \pm \infty$ and $\bar\xi_1 \ra \mp \infty$ as $\sigma t\ra \pm \infty$. Then, by taking the limit of of solution~\eref{2solution} when $\xi_1,-\bar\xi_1\ra \infty$ and $\xi_1,-\bar\xi_1\ra -\infty$, we obtain another asymptotic expression in the rational form
\begin{subequations}
\begin{align}
&u\ra  u_{4}^{\pm}=\rho  e^{2\, i \rho^2 t + i \phi}\bigg[1+\frac{2\,\sigma }{2\, i \rho  (\bar{\eta}_1+\gamma^*_2)+\sigma \pm 2\, i \sec(\theta)}\bigg] \quad (\sigma\,t \ra  \pm \infty),\label{4asymptotica}\\
&|u| \ra  |u_{4}^{\pm}|=\rho^2\bigg\{1+ \frac{8\,(1+\rho\, \sigma \gamma_{2I}) }{4\big[-\rho \bar\eta_1-\rho \gamma_{2R}\mp \sec(\theta) \big]^2+(2\rho \gamma_{2I}+\sigma)^2}\bigg\}^{\frac{1}{2}}. \label{4asymptoticb}
\end{align}
\end{subequations}
When condition~\eref{cond2} is satisfied, $u_{4}^{\pm}$  is also nonsingular and it can represent an RAD soliton  for $\sigma \gamma_{2I} > -\frac{1} {\rho}$ or an RD soliton   for $\sigma \gamma_{2I}<-\frac{1}{\rho}$ in the line $\bar\eta_1+ \gamma_{2R} \pm \frac{\sec(\theta)}{ \rho}=0 $.
\end{enumerate}

\subsection{Properties of soliton interactions}

As shown in the above asymptotic analysis, solution~\eref{2solution} admits four pairs of asymptotic solitons  $(u^+_i, u^-_i)$ ($1\leq i\leq 4$)
which are localized in four different directions in the $xt$ plane. Below, based on  Eqs.~\eref{1asymptotica}--\eref{4asymptoticb},
we further reveal the soliton interaction properties described by  solution~\eref{2solution}:
\begin{enumerate}
\item[(i)]
For each pair of asymptotic solitons $(u^+_i, u^-_i)$, their intensities have the same amplitudes (i.e., $|u^\pm_i|_{\rm{max}}^2 -\rho^2 $ for the antidark soliton or  $\rho^2 -|u^\pm_i|_{\rm{min}}^2  $ for the dark soliton):
\begin{subequations}
\begin{align}
&A_1^\pm = \frac{2\rho^2\sqrt{1-b^2} |\sin(\varphi_1)| }{1- \sgn(b) \sin (\theta+\varphi _1)
},    \quad  A_2^\pm=   \frac{2\rho^2\sqrt{1-b^2}\, |\sin (2\theta +\varphi _1)|}{1-\sgn(b)\sin(\theta +\varphi _1)  },   \label{2zf}  \\
& A_3^\pm=\frac{8\rho^3 | \gamma_{2I}| }{(1+2 \rho \sigma \gamma_{2I})^2},  \qquad\qquad\,\,\,\,
A_4^\pm=\frac{8 \rho^2 |1+ \sigma\rho\gamma_{2I}|}{(1+2 \sigma \rho\gamma_{2I})^2}.     \label{34zf}
\end{align}
\end{subequations}

\item[(ii)]
The envelope velocity of $u^+_i$ is  exactly equal to that
of $u^-_i$, i.e.,
\begin{align}
v_1^{+}=v_1^{-}=-2\,\rho \,b, \quad v_2^{+}=v_2^{-}=2\,\rho\,b,\quad v^+_3=v^-_3=2\,\sigma\rho, \quad v_4^{+}=v^-_4=-2\,\sigma\rho.
\end{align}
\item[(iii)]
All the exponential and rational solitons undergo the phase shifts for their envelopes and the phase differences can be  given by
\begin{align}
\delta\phi_1=-\delta\phi_2=\Delta_1^{+}-\Delta_1^{-}= -2\ln |b|, \quad
\delta\phi_3=-\delta\phi_4= 2\sec(\theta), \label{delphi1}
\end{align}
which is  contrast to that there is no phase shift in the  rational soliton solutions~\cite{LiXu1}.
\end{enumerate}
Therefore,  all the  interacting solitons can retain their individual shapes, amplitudes and velocities upon their mutual interactions except for some phase shift, which meets the nature of  elastic soliton interactions.

\begin{figure}[ht]
 \centering
\subfigure[]{\label{Fig1a}
\includegraphics[width=1.7in]{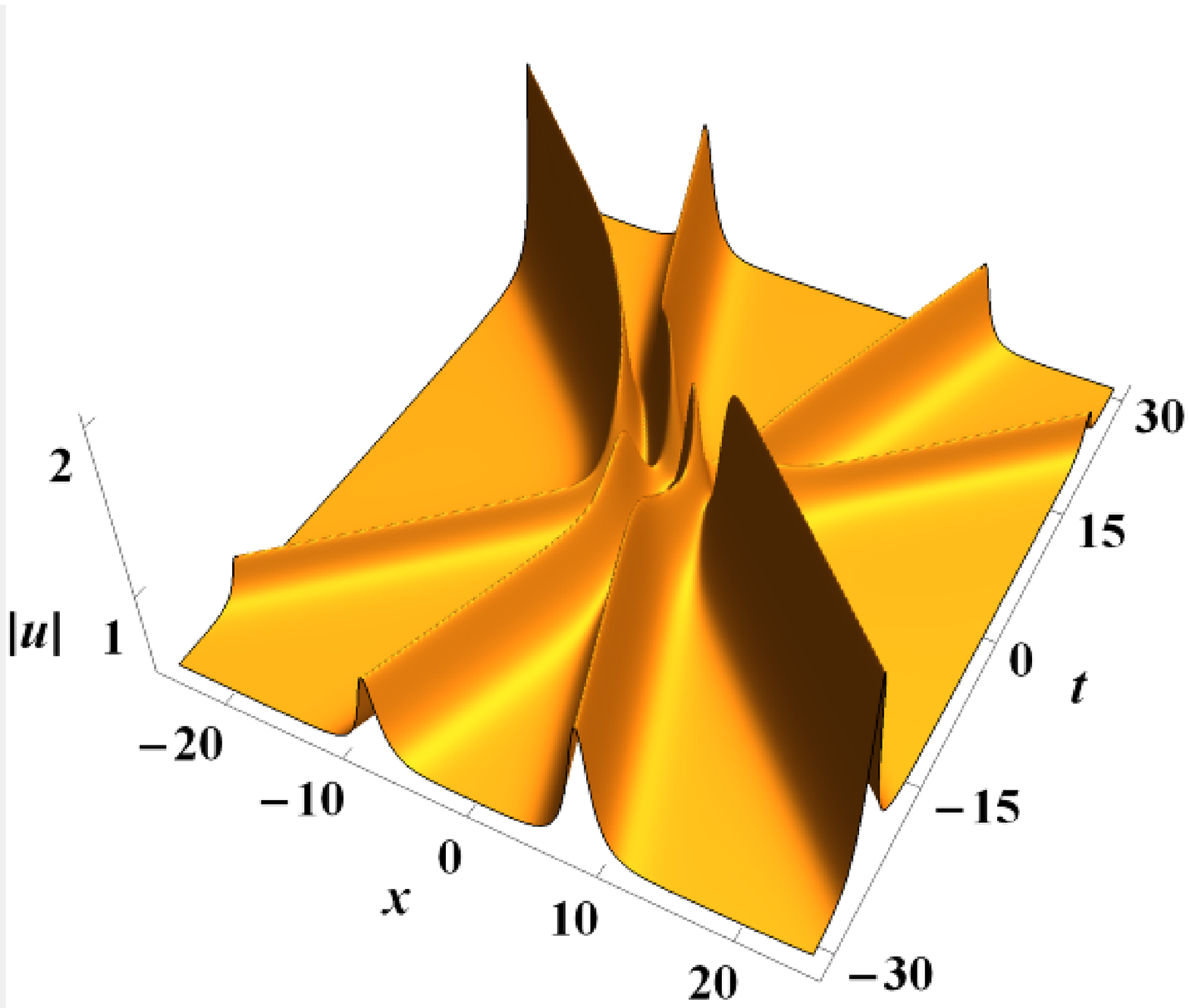}}\hfill
\subfigure[]{ \label{Fig1b}
\includegraphics[width=1.7in]{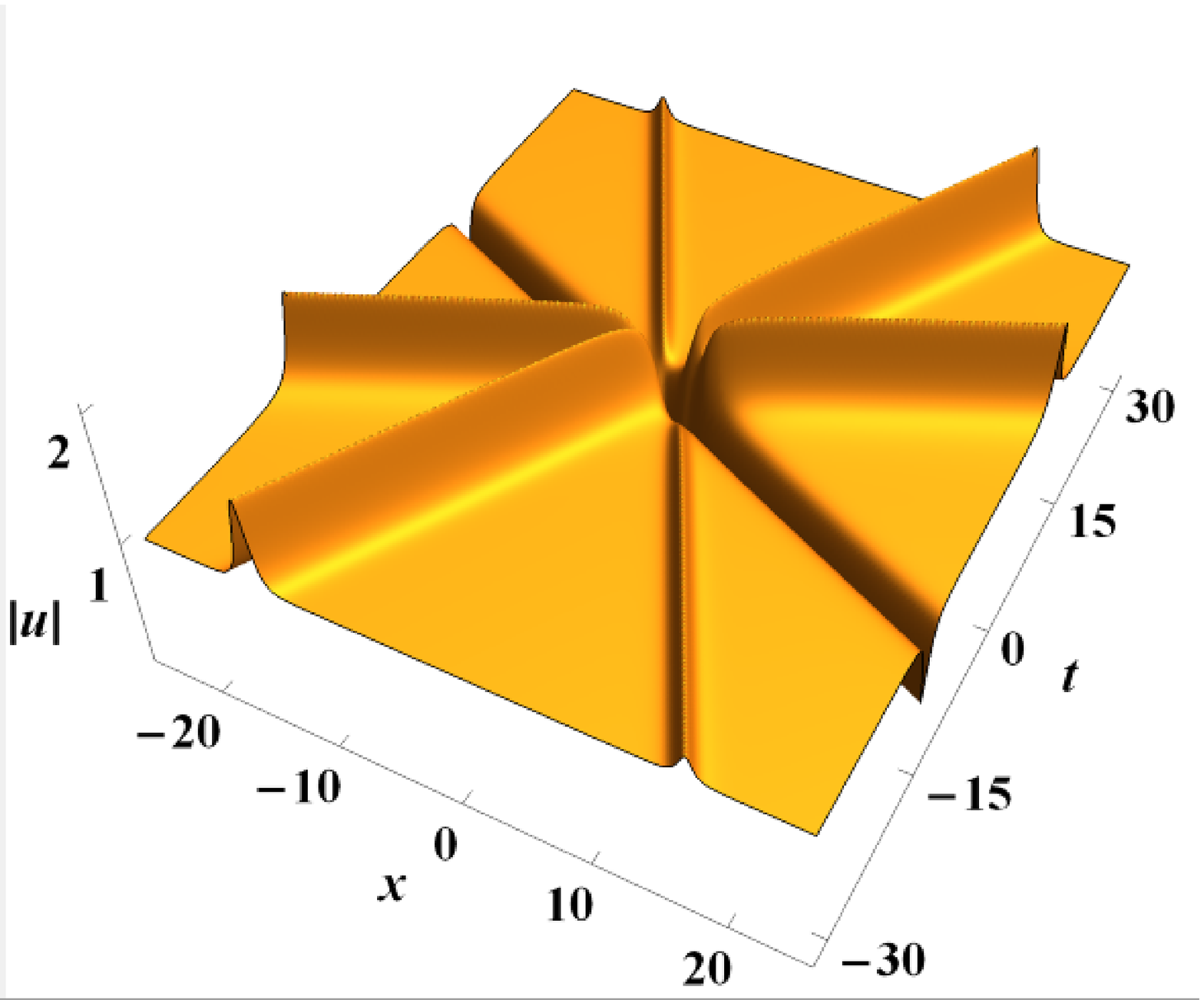}}\hfill
\subfigure[]{ \label{Fig1c}
\includegraphics[width=1.7in]{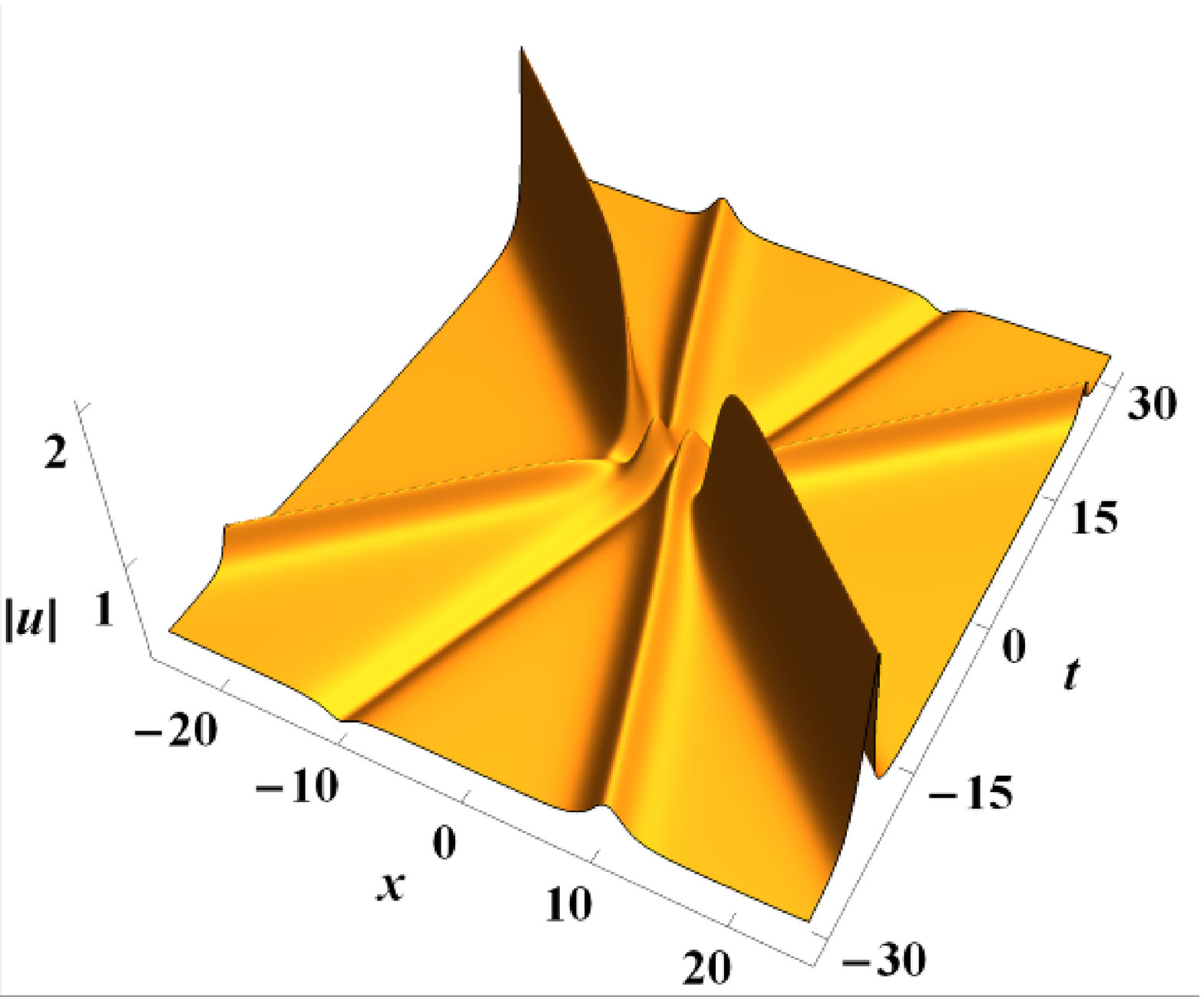}}\\
\subfigure[]{\label{Fig1d}
\includegraphics[width=1.7in]{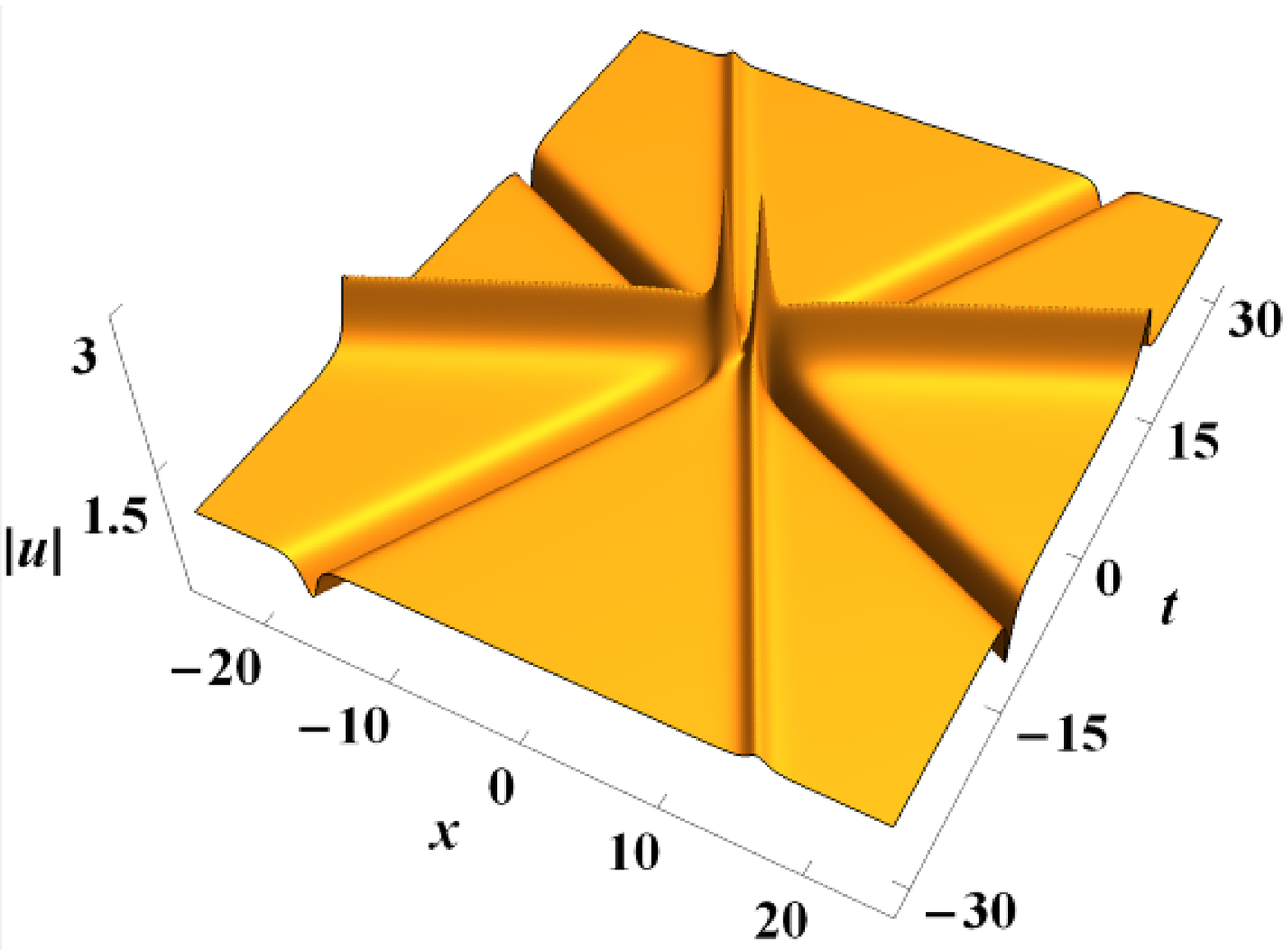}}\hfill
\subfigure[]{ \label{Fig1e}
\includegraphics[width=1.7in]{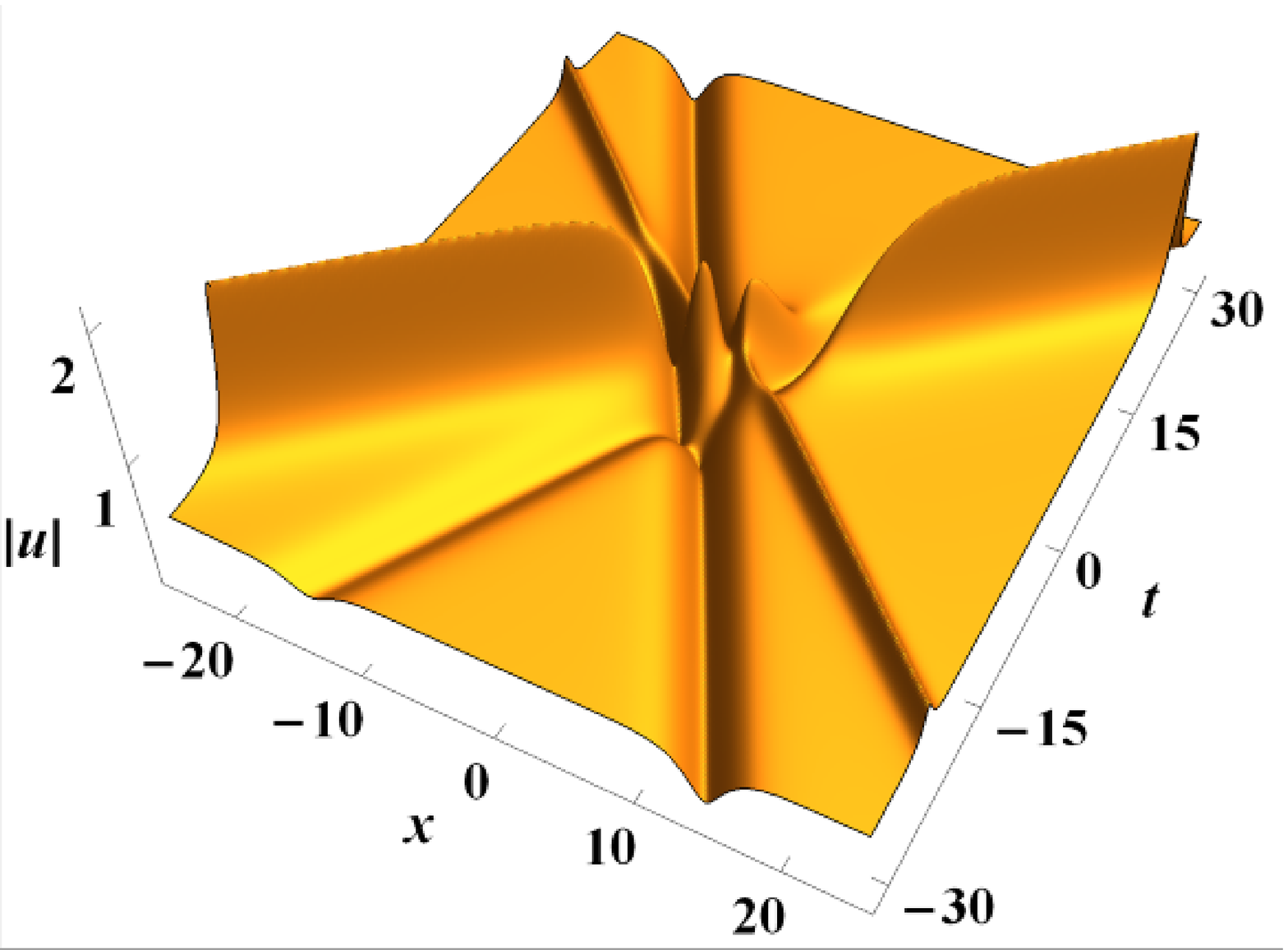}}\hfill
\subfigure[]{ \label{Fig1f}
\includegraphics[width=1.7in]{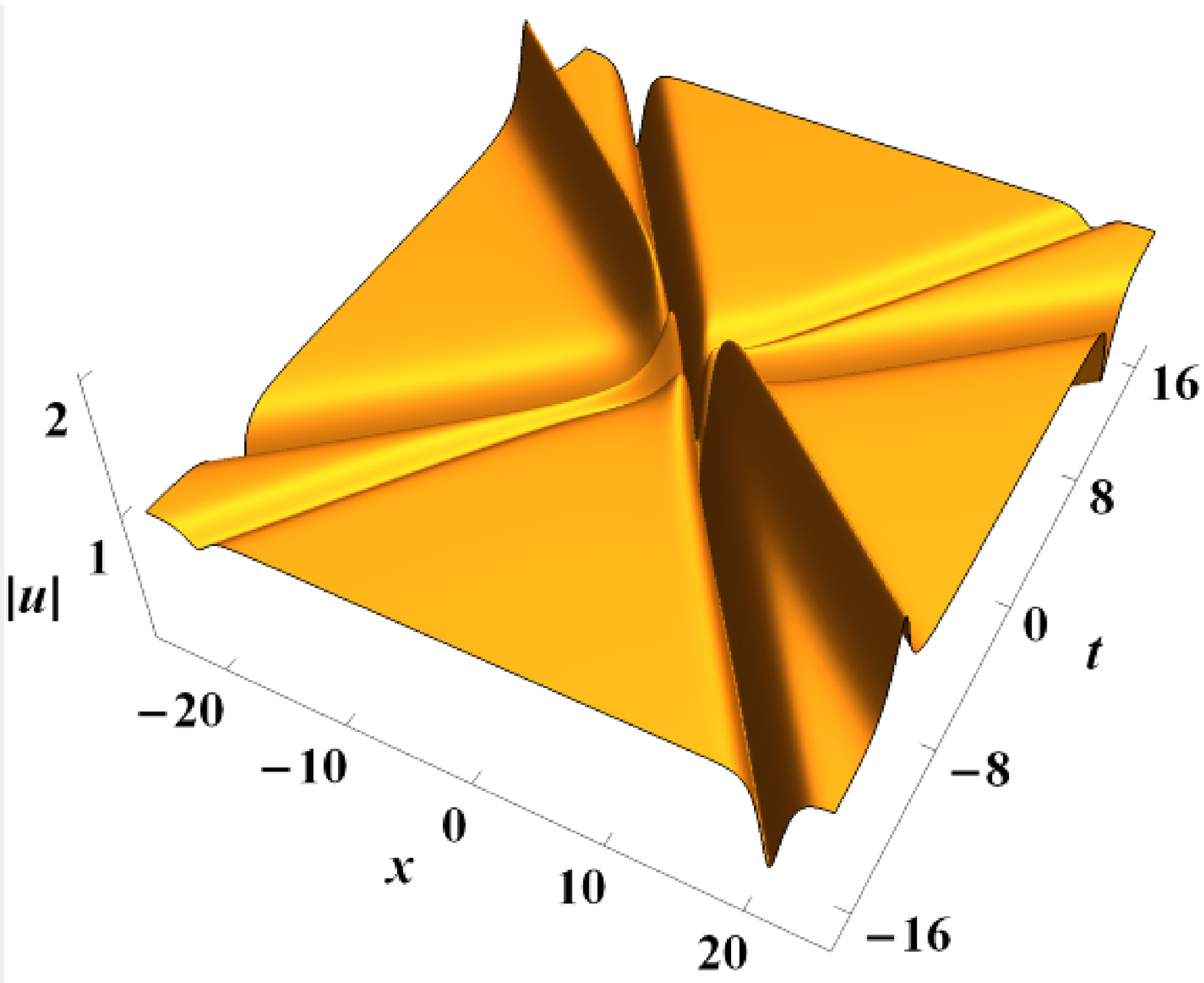}}\hfill
\caption{\small Six types of non-degenerate four-soliton interactions via
solution~\eref{2solution}:  (a) EAD-EAD-RAD-RAD soliton interaction with $\rho=\frac{1}{2}$, $b=\frac{3}{10}$, $\phi=0$, $\sigma=-1$, $\gamma_1=-\frac{1}{20} + \frac{1}{10}\, i $  and $\gamma_2=\frac{1}{5}\, i $; (b)  EAD-EAD-RD-RAD soliton interaction with $\rho=1$, $b=\frac{1}{4}$, $\phi=0$, $\sigma=1$, $\gamma_1=1+\frac{1}{8}\, i $  and $\gamma_2= i $; (c) EAD-ED-RAD-RAD soliton interaction with $\rho=\frac{1}{2}$,  $b=\frac{3}{10}$, $\phi=0$, $\sigma=-1$,  $\gamma_1=-1+\frac{1}{2}\, i $  and $\gamma_2=\frac{1}{5}\, i $; (d) EAD-ED-RD-RAD soliton interaction with $\rho=1$, $b=\frac{1}{4}$, $\phi=0$, $\sigma=1$,  $\gamma_1=-1+\frac{1}{8}\, i $  and $\gamma_2= i $; (e) ED-ED-RAD-RAD soliton interaction with $\rho=\frac{1}{2}$, $b=\frac{1}{2}$, $\phi=0$, $\sigma=1$,  $\gamma_1=-\frac{1}{5}\, i $  and $\gamma_2=-\frac{1}{5}\, i $; (f) ED-ED-RAD-RD soliton interaction with $\rho=1$, $b=\frac{3}{5}$, $\phi=0$, $\sigma=1$,  $\gamma_1=-2\, i $  and $\gamma_2=-2\, i $.  \label{Fig1} }
\end{figure}

\begin{figure}[ht]
 \centering
\subfigure[]{\label{Fig2a}
\includegraphics[width=1.8in]{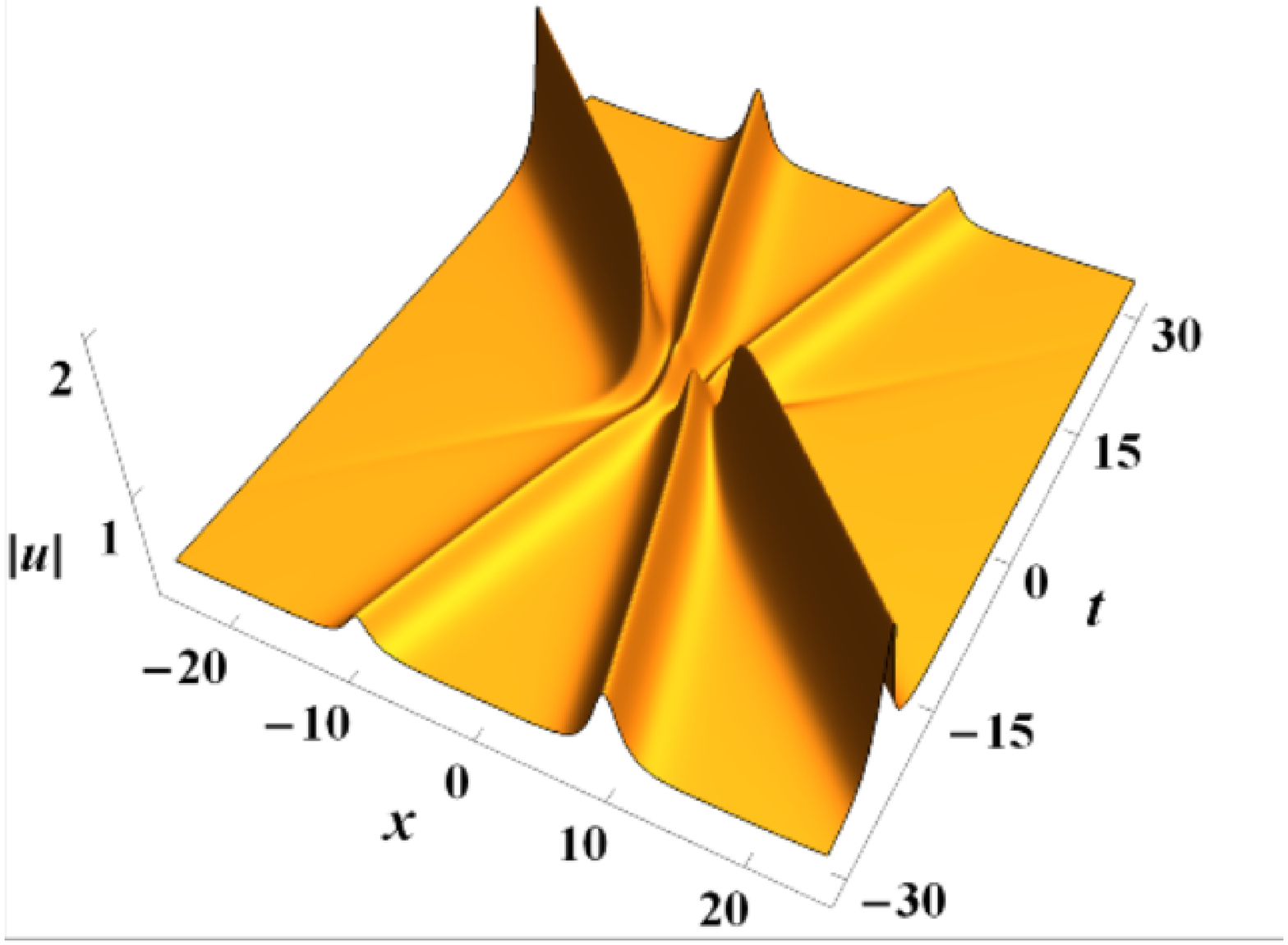}}\hfill
\subfigure[]{ \label{Fig2b}
\includegraphics[width=1.8in]{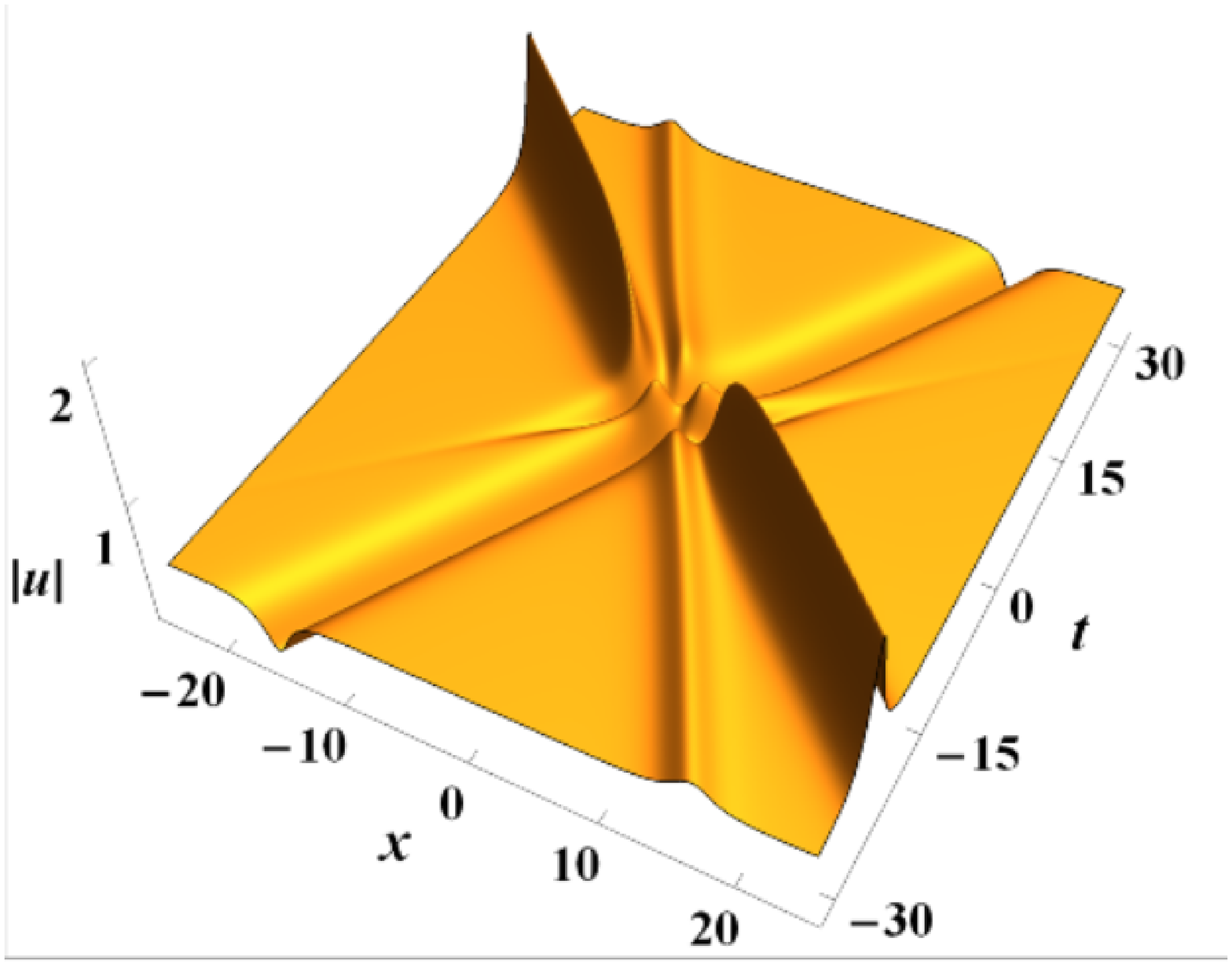}}\hfill
\subfigure[]{ \label{Fig2c}
\includegraphics[width=1.8in]{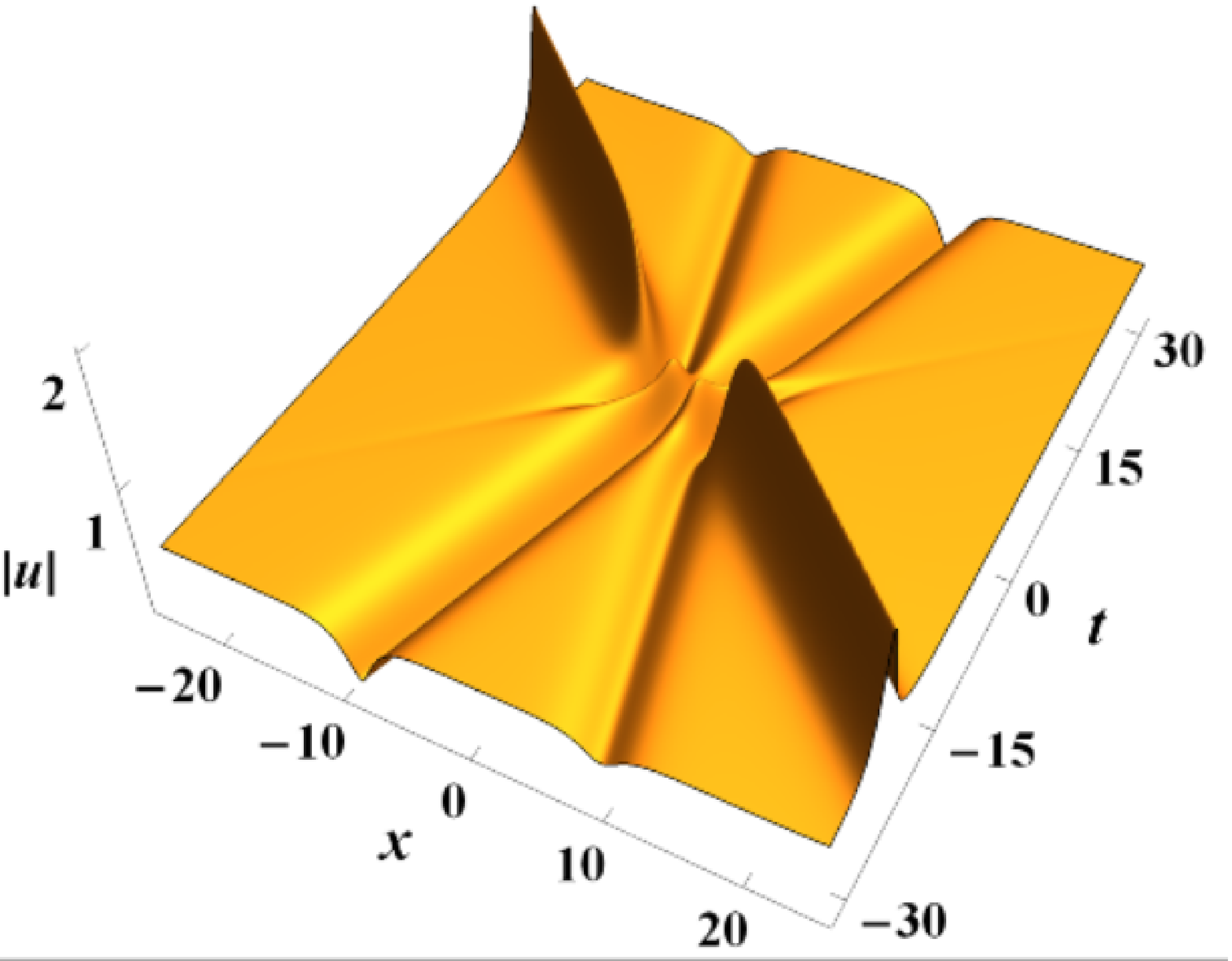}}\\
\subfigure[]{\label{Fig2d}
\includegraphics[width=1.8in]{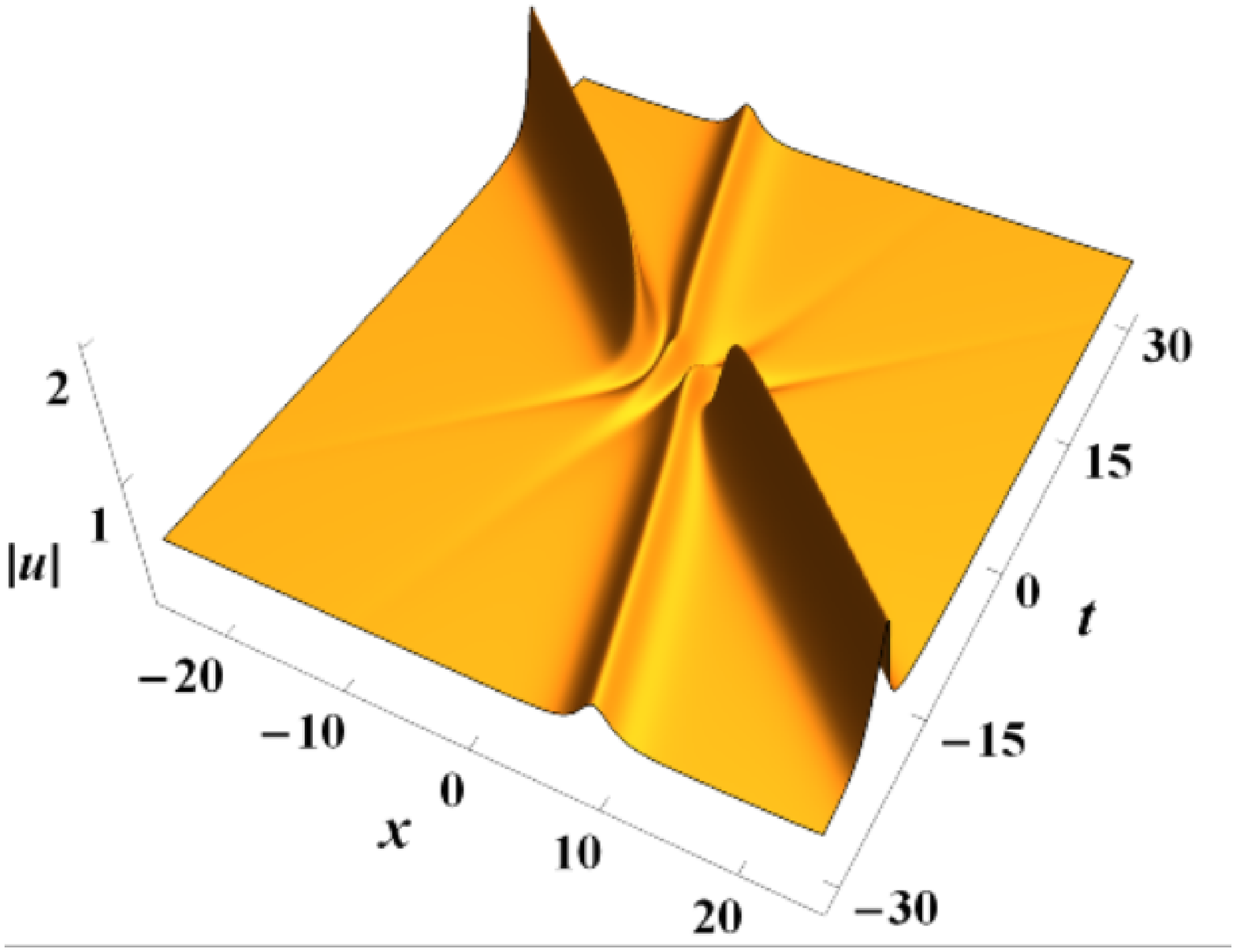}}\hspace{2.5cm}
\subfigure[]{ \label{Fig2e}
\includegraphics[width=1.8in]{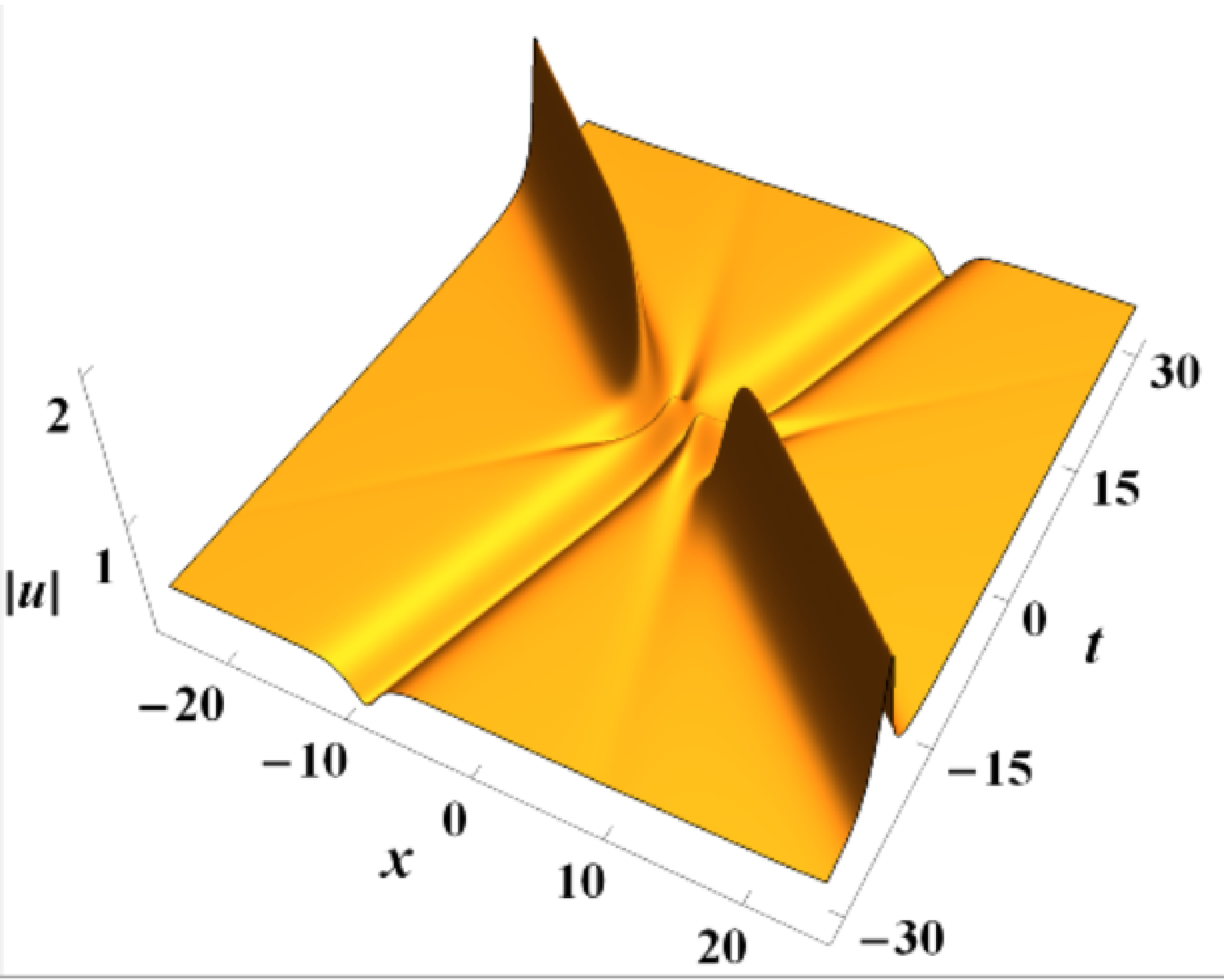}}\hfill
\caption{\small Five types of degenerate four-soliton interactions via
solution~\eref{2solution}:  (a) EAD-EAD-V-RAD soliton interaction with $\rho=\frac{1}{2}$, $b=\frac{1}{4}$, $\phi=0$, $\sigma=-1$, $\gamma_1=-1+ i $ and $\gamma_2=0$. (b)  EAD-ED-V-RAD  soliton interaction with $\rho=\frac{1}{2}$, $b=\frac{1}{2}$, $\phi=0$, $\sigma=-1$, $\gamma_1=-1+ \frac{1}{2} i $  and $\gamma_2=0$. (c) ED-ED-V-RAD soliton interaction with $\rho=\frac{1}{2}$, $b=\frac{1}{4}$, $\phi=0$, $\sigma=-1$,  $\gamma_1=-1-\frac{1}{2}\, i $ and $\gamma_2=0$. (d) EAD-V-V-RAD soliton interaction with $\rho=\frac{1}{2}$, $b=\frac{1}{4}$, $\phi=0$, $\sigma=-1$, $\gamma_1=-\frac{7}{2\sqrt{15}}+\frac{1}{2}\, i $ and $\gamma_2=0$. (e) V-ED-V-RAD soliton interaction with $\rho=\frac{1}{2}$, $b=\frac{1}{4}$, $\phi=0$, $\sigma=-1$,  $\gamma_1=-1$ and $\gamma_2=0$. Here, ``V'' represents the vanishment of an asymptotic soliton as $|t|\ra \infty$. \label{Fig2}}
\end{figure}

Recalling that each pair of asymptotic solitons $(u^+_i, u^-_i)$ ($1\leq i\leq 4$) in solution~\eref{2solution} could be of the dark or antidark type, one can obtain a large variety of elastic soliton interactions. To illustrate,  Figs.~\ref{Fig1a}--\ref{Fig1f} present some examples of four-soliton interactions. In particular, some soliton pair(s) $(u^+_i, u^-_i)$  will vanish (i.e., the  amplitude of $|u^\pm_i|^2$ becomes zero), and the relevant parametric condition is  $\sin \left(\varphi _1\right) =0$ for $i=1$, $\sin \left(2\theta +\varphi _1\right) =0$  for $i=2$, $ \gamma_{2I}=0$ for $i=3$, and $\sigma  \gamma_{2I}=-\frac{1}{\rho}$ for $i=4$.
For those particular cases, the four-soliton interaction degenerates to a three-soliton interaction or even to a two-soliton interaction, as shown in Figs.~\ref{Fig2a}--\ref{Fig2e}. However, the degenerate four-soliton interactions cannot be simply regarded as the conventional three- or two-soliton interactions since there are still the trace for the vanishing asymptotic soliton(s) in the near-field region (see Figs.~\ref{Fig2a}--\ref{Fig2e}). In Tables~\ref{Table1} and~\ref{Table2}, we list all the possible cases of the exponential and rational asymptotic solitons and their related  parametric conditions. The combinatorial calculation indicates that solution~\eref{2solution} can describe a total of forty different types of soliton interactions.

\begin{table}[h]
\small
\caption{Types of the exponential asymptotic solitons $u^\pm_1$ and $u^\pm_2$ with
different parametric conditions. \label{Table1}}
\begin{center}
\begin{tabular}{|c|c|c|c|c|}
\hline  \multicolumn{1}{|c|}{Parametric conditions} &   \tabincell{c}{Asymptotic
soliton $u_{1}^{\pm}$  \\  $[\sgn(b)\,t \ra \pm \infty]$} & \tabincell{c}{Asymptotic soliton $u_{2}^{\pm}$ \\ $[\sgn(b)\,t \ra \pm \infty]$}
\\  \hline
$ b \sin \left(\varphi _1\right) >0$,  $b \sin \left(2\theta +\varphi _1\right) >0$   &  EAD soliton  &   EAD soliton
\\  \hline
$ b \sin \left(\varphi _1\right) >0$,  $b \sin \left(2\theta +\varphi _1\right) <0$    &  EAD soliton  &   ED soliton
\\  \hline
$ b \sin \left(\varphi _1\right) <0$,  $b \sin \left(2\theta +\varphi _1\right) >0$   &  ED soliton  &   EAD soliton
\\  \hline
$ b \sin \left(\varphi _1\right) <0$,  $b \sin \left(2\theta +\varphi _1\right) <0$   &  ED soliton  &   ED soliton
\\ \hline
$ b \sin \left(\varphi _1\right) >0$,  $\sin \left(2\theta +\varphi _1\right) =0$   &  EAD soliton  &   Vanish
\\  \hline
$ b \sin \left(\varphi _1\right) <0$,  $\sin \left(2\theta +\varphi _1\right) =0$   &  ED soliton  &   Vanish
\\  \hline
$ \sin \left(\varphi _1\right) =0$, $b \sin \left(2\theta +\varphi _1\right) >0$   &  Vanish  &   EAD soliton
\\  \hline
$\sin \left(\varphi _1\right) =0$, $b \sin \left(2\theta +\varphi _1\right) <0$    &  Vanish  &   ED soliton
\\  \hline
\end{tabular}
\end{center}
\end{table}
\begin{table}[h]
\small
\caption{Types of the rational asymptotic solitons $u^\pm_3$ and $u^\pm_4$ with
different parametric conditions. \label{Table2}}
\begin{center}
\begin{tabular}{|c|c|c|c|c|}
\hline  \multicolumn{1}{|c|}{Parametric conditions} & \tabincell{c}{Asymptotic soliton $u_{3}^{\pm}$ \\ ($\sigma\,t \ra \pm \infty$)}& \tabincell{c}{Asymptotic soliton $u_{4}^{\pm}$ \\ ($\sigma\,t \ra \pm \infty$) }
\\ \hline
$\sigma  \gamma_{2I}<-\frac{1}{\rho}$   &  RAD soliton  &  RD soliton
\\ \hline
$  -\frac{1}{\rho}<\sigma \gamma_{2I}<0$   &  RAD soliton  &  RAD soliton
\\ \hline
$ \sigma  \gamma_{2I}>0$   &  RD soliton   &  RAD soliton
\\ \hline
$   \gamma_{2I}=0$     &  Vanish   &  RAD soliton
\\ \hline
$\sigma  \gamma_{2I}=-\frac{1}{\rho}$   &  RAD soliton  &  Vanish
\\ \hline
\end{tabular}
\end{center}
\end{table}

\section{The second type of mixed soliton solution}
\label{Sec4}

In this section, we consider that both $ \lam_1$ and $\lam_2$ degenerate to $\mi \,b \rho $ with $0 < \mid b \mid < 1  $.  For simplicity, we set  $\epsilon_1=\epsilon_2=\delta^2$ and $ \lam_1= \lam_2 = \mi \, \sigma  \rho(1+\delta^2) $, and take
\begin{align}
\alpha_{1,2} = \alpha'_1 e^{h_1(s_1 + s_2 \delta^2)}, \quad \beta_{1,2}= \beta'_1 e^{-h_1(s_1 + s_2 \delta^2)},
\end{align}
where $|\delta|\ll 1$ is a small parameter, and $s_1$ and $s_2$ are two arbitrary complex numbers. Based on the generalized DT, we must expand $f_k$ and $g_k$ ($k=1,2$) at $\lam_{1,2} = i\,b\rho $ in the way of Eqs.~\eref{Taylor1} and~\eref{Taylor2} up to $j=1$. Then,   the second type of mixed soliton solution can be obtained as follows:
\begin{align}
\hspace{-10mm}  u = \rho\, e^{2\, i \rho^2 t + i \phi} + 2\,\frac{\begin{vmatrix}
f_1^{(0,0)} & f_1^{(1,0)} & f_1^{(2,0)} &  g_1^{(0,0)}  \\
f_1^{(0,1)} & f_1^{(1,1)} & f_1^{(2,1)} &  g_1^{(0,1)} \\
- \bar{g}_1^{(0,0)} &  - \bar{g}_1^{(1,0)} &  - \bar{g}_1^{(2,0)} & \bar{f}_1^{(0,0)}   \\
- \bar{g}_1^{(0,1)} &  - \bar{g}_1^{(1,1)}  &  - \bar{g}_1^{(2,1)} & \bar{f}_1^{(0,1)}
\end{vmatrix}}{
\begin{vmatrix}
f_1^{(0,0)} & f_1^{(1,0)} &  g_1^{(0,0)} &  -g_1^{(1,0)} \\
f_1^{(0,1)} & f_1^{(1,1)} &  g_1^{(0,1)} &  -g_1^{(1,1)} \\
- \bar{g}_1^{(0,0)} &  - \bar{g}_1^{(1,0)} & \bar{f}_1^{(0,0)} & -\bar{f}_1^{(1,0)}  \\
- \bar{g}_1^{(0,1)} &  - \bar{g}_1^{(1,1)} & \bar{f}_1^{(0,1)} & -\bar{f}_1^{(1,1)}
\end{vmatrix}}, \label{3solution}
\end{align}
with
\begin{subequations}
\begin{align}
& f_1^{(0,0)}=\alpha'_1 e^{ i  \rho^2 t +\frac{ i  \phi }{2}}\big(e^{\kappa (\xi_1+s_1)} + \gamma_1 e^{-\kappa (\xi_1+s_1)}\big), \label{Eq30a}  \\
& f_1^{(0,1)}= - \frac{\alpha'_1 \rho \chi(x,t)  }{\sqrt{1-b^2}} e^{ i  \rho^2 t +\frac{ i  \phi }{2}} \big(e^{\kappa (\xi_1+s_1)} - \gamma_1 e^{-\kappa (\xi_1+s_1)} \big),   \label{Eq30b} \\
&f_1^{(k,0)}=( i \rho b)^k f_1^{(0,0)}, \quad  f_1^{(k,1)}=( i \rho b)^k f_1^{(0,1)}+k f_1^{(k,0)}\quad (k=1,2),  \label{Eq30c} \\
& g_1^{(0,0)}= \alpha'_1 e^{- i  \rho^2 t -\frac{ i  \phi }{2}}\big(e^{\kappa (\xi_1+s_1)- i \theta} - \gamma_1  e^{-\kappa (\xi_1+s_1) + i \theta} \big),  \label{Eq30d}   \\
& g_1^{(0,1)}=\frac{- \alpha'_1[\rho\chi(x,t)+ i b] }{\sqrt{1-b^2}} e^{- i  \rho^2 t -\frac{ i  \phi }{2}}\big(e^{\kappa (\xi_1+s_1)- i \theta} + \gamma_1  e^{-\kappa (\xi_1+s_1)+i \theta} \big),  \label{Eq30e}    \\
&g_1^{(k,0)}=(i\rho b)^k g_1^{(0,0)}, \quad  g_1^{(k,1)}=( i \rho b)^k g_1^{(0,1)}+k g_1^{(k,0)}\quad (k=1,2),  \label{Eq30f} \\
& \chi(x,t)=(b^2-1)(2 b \rho t+s_2)+b^2 (\xi_1+s_1), \quad \xi_1=x+2 b \rho t,  \label{Eq30g} \\
& \kappa=\rho\sqrt{1-b^2}, \quad \theta=\arctan(b/\sqrt{1-b^2}),\quad \gamma_1=\beta'_1/\alpha'_1=r_1  e^{ i \varphi_1}, \label{Eq30h}
\end{align}
\end{subequations}
where $-\pi/2<\theta<\pi/2$, $-\pi<\varphi_1<\pi $, $r_1>0$ is a real constant, the $bar$ denotes the combination of complex conjugate and space reversal, and $\alpha'_1$ will be canceled out when Eqs.~\eref{Eq30a}--\eref{Eq30h} are substituted into Eq.~\eref{3solution}.
Also, this solution is in the mixed exponential-rational form since it includes both the exponential terms $e^{\pm\kappa\xi_1}, e^{\pm\kappa\bar\xi_1}$  and  algebraic terms $\chi(x,t), \bar\chi(x,t)$. In the following, we will develop the asymptotic analysis method so as to understand the solitonic behavior in solution~\eref{3solution}.

\subsection{Asymptotic analysis}\label{sec4.1}

To begin with, we argue that  the asymptotic solitons of solution~\eref{3solution} cannot be located in any straight line $x- c\,t'=\textit{const}$ with $t'=\sgn(b)t$. Noticing that  $\xi_1-(x- c\,t')=( c+ 2 |b| \rho )t'$ and $-\bar\xi_1-(x- c\,t')=( c- 2 |b| \rho )t'$,  we have
the asymptotic behavior of  $\xi_1$ and $-\bar\xi_1$ when $t'\ra \pm \infty$:
\begin{align}
\xi_1 \ra \left\{
\begin{array}l
\pm \infty,  \, \,\,\,\,\,  c > -2 |b| \rho,\\
O(1),  \, \,\,\,\, c = -2 |b| \rho, \\
\mp \infty,  \, \,\,\,\,\, c < -2 |b| \rho,
\end{array}
\right. \quad\,\,
\bar\xi_1  \ra \left\{
\begin{array}l
\mp \infty,  \, \,\,\,\,\,  c > 2 |b| \rho,\\
O(1),  \, \,\,\,\, c = 2 |b| \rho, \\
\pm \infty,  \, \,\,\,\,\, c < 2 |b| \rho.
\end{array}
\right.  \label{Eq50}
\end{align}
Thus, no matter whether  $c$ is equal to $\pm2 |b| \rho$ or not, the limit of solution~\eref{3solution}  along the line $x-c\,t'=\textit{const}$ as $t'\ra \pm \infty$ is a plane wave, that is,
\begin{align}
\displaystyle\lim_{t'\ra \pm\infty} u
=\begin{cases}\rho  e^{2\, i \rho^2 t +  i \phi}\quad\quad\quad\, (|c|>2|b|\rho),\\
\rho  e^{2\, i \rho^2 t +  i \phi \pm 4 i \theta}\quad\,\,\,\, (-2|b|\rho<  c< 2|b|\rho), \\
-\rho  e^{2\, i \rho^2 t +  i \phi \pm 2 i \theta }\quad(c= \pm2 |b| \rho ),
\end{cases} \label{ThreePW}
\end{align}
which implies that there is no asymptotic soliton lying in any straight line of the $xt$ plane.

Next, we consider that the asymptotic solitons of solution~\eref{3solution} are located in some curves $F(x,t')=0$. Because $-\frac{F_{t'}}{F_x}\neq \textit{const}$, both $\xi_1$ and $\bar\xi_1$ will tend to $+\infty$ or $-\infty$ along $F(x, t')=0$ as $t'\ra \infty$. Thus, before explicitly determining the curves $F(x,t')=0$, one can calculate the \emph{intermediate} asymptotic expressions of solution~\eref{3solution} by letting $\bar\xi_1\ra\pm\infty$ or $\xi_1\ra\pm\infty$. If precedently taking $\bar\xi_1\ra\pm\infty$ for solution~\eref{3solution}, we have the limits as follows:
\begin{align}
 u\ra  u_{\rm{I}}=& \rho  e^{2\, i \rho^2 t +  i \phi}  \notag \\
& \hspace{-1mm} \times \! \bigg\{1+\frac{4 b \sqrt{1-b^2}[ e^{2\kappa(\xi_1 + s_1)} + \gamma_1 (1+\sqrt{1-b^2}\,\zeta)]} {\omega e^{2\kappa(\xi_1 + s_1)}-2 b \sqrt{1-b^2}\, \gamma_1 (1+\zeta e^{- i \theta}) - i b^4 \gamma^2_1  e^{-2\kappa(\xi_1+s_1)} }\bigg\}\quad (\bar\xi_1\ra +\infty),  \label{3formsa} \\[1mm]
u\ra  u_{\rm{I\!I}}=&\rho  e^{2\, i \rho^2 t + i \phi} \notag \\
& \hspace{-1mm} \times \! \bigg\{1+\frac{4b\sqrt{1-b^2}\,[
\gamma_1  e^{2\kappa(\xi_1 + s_1)}(1-\sqrt{1-b^2}\,\zeta)+\gamma^2_1]}
{ i  b^4 e^{4\kappa(\xi_1 + s_1)}-2 b \sqrt{1-b^2}\,\gamma_1  e^{2\kappa(\xi_1 + s_1)}(1-\zeta e^{ i \theta})+\omega^{*}\gamma^2_1 }\bigg\}\quad (\bar\xi_1\ra -\infty), \label{3formsb}
\end{align}
with $\omega=- i  e^{-2 i \theta}$ and  $\zeta = 2 \rho[b^2(\xi_1+s_1+s_2)-s_2] -4 \rho^2 |b|(1-b^2)t' $. Here, both Eqs.~\eref{3formsa} and~\eref{3formsb} contain two independent variables $\xi_1$ and $t'$, so that the soliton center trajectories cannot be directly obtained by calculating the extreme values of $|u_{\rm{I}}|^2$ and $|u_{\rm{I\!I}}|^2$.

\begin{figure}[ht]
 \centering
\includegraphics[width=2.8in]{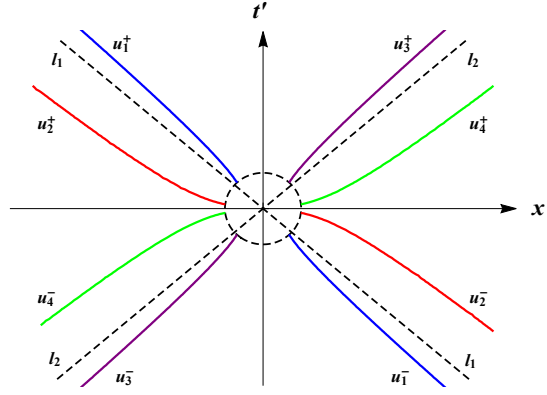}
\caption{\small Schematic diagrams for the distribution of eight asymptotic solitons in solution~\eref{3solution}, where the dashed oval stands for the soliton interaction region. \label{Fig4}}
\end{figure}

In fact, there must be some  balance between $\xi_1$ and $t'$ in Eqs.~\eref{3formsa} and~\eref{3formsb} when an asymptotic soliton appears as $t'\ra \pm \infty$.  By the method of dominant balance~\cite{AMMSE}, we assume that
\begin{align}
t' \sim W(x,t) ( e^{2\kappa\xi_1})^p, \quad W(x,t) = O(1), \label{DRE}
\end{align}
where $p$ is a constant to be determined. It can be found that as $t'\ra \pm \infty$ both Eqs.~\eref{3formsa} and~\eref{3formsb}
approaches a plane wave as given in Eq.~\eref{ThreePW} for all the cases when $p\neq \pm1$. Therefore, Eq.~\eref{DRE} with
$p = \pm 1 $ is the only allowed balance for deriving the asymptotic solitons from Eqs.~\eref{3formsa} and~\eref{3formsb}.
With an elaborate computation on Mathematica, we obtain that there are four asymptotic solitons in the mixed
exponential-rational form: 
\begin{enumerate}
\item[(i)]
If  $ W(x,t) =t'  e^{-2\kappa\xi_1} = O(1)$, we take the limit of $u_{\rm{I}}$ when $\xi_1\ra +\infty $, yielding
\begin{subequations}
\begin{align}
& u_{\rm{I}} \ra u^+_{1}=\rho  e^{2\, i \rho^2 t + i \phi}
\bigg\{1+ \dfrac{4 b \sqrt{1-b^2} \omega^{*}\big[ 1-4\rho^2 \gamma_1 b(1-b^2)^{\!\frac{3}{2}} t\,e^{-2\kappa(\xi_1+s_1) }\big] }{1 +8\, i \rho^2 \gamma_1 b^2  (1-b^2)^{\!\frac{3}{2}}  e^{ i \theta}t\,  e^{-2\kappa (\xi_1+s_1)}  }\bigg\},
\label{1finala} \\
& |u_{\rm{I}}|\ra |u^+_{1}|=\rho\bigg[1+\dfrac{2 \sqrt{1-b^2} \sin(\varphi_1-2\kappa s_{1I})}
{\sgn(b) \cosh(2\kappa\xi_1+ 2\kappa s_{1R}-\ln \Xi_1^+)-\sin(\theta+\varphi_1-2\kappa s_{1I})}\bigg]^{\frac{1}{2}}, \label{1finalb}
\end{align}
\end{subequations}
with  $\Xi_1^+=8 \rho^2 r_1 b^2 (1-b^2)^{\!\frac{3}{2}} t'$. When $\varphi_1$, $b$ and $s_1$ satisfy
\begin{align}
\theta+\varphi_1-2\kappa s_{1I}\neq2k\pi+\frac{1}{2}\sgn(b)\pi \quad  (k \in \mathbb{Z}), \label{conda}
\end{align}
$u^+_{1}$ is nonsingular and it can represent an mixed antidark (MAD) soliton for $\sgn(b)\sin(\varphi_1-2\kappa s_{1I})>0$ or an mixed dark (MD) soliton for $\sgn(b)\sin(\varphi_1-2\kappa s_{1I})<0$. Note that the asymptotic expression~\eref{1finala} is obtained by orderly letting $ \bar\xi_1,\xi_1 \ra +\infty$. Since $\xi_1+\bar\xi_1 =4 \rho |b| t'\ra +\infty$, $u^+_{1}$ appears as an asymptotic state of solution~\eref{3solution} as $t'\ra +\infty$. Meanwhile, calculating the extreme value of $|u^+_{1}|^2$ shows that the soliton center trajectory is
\begin{align}
\mathcal{C}^+_{1}:\,\,t'  e^{-\kappa\xi_1} =\frac{ e^{2\kappa s_{1R}} }{8\rho^2 r_1 b^2(1-b^2)^{\!\frac{3}{2}}}\quad (t' >0), \label{Traj1}
\end{align}
where its slope is  given by
\begin{align}
K^+_{1}=-\frac{1}{2 |b| \rho}+\frac{1}{2 |b| \rho -8 \rho^3 b^2 \sqrt{1-b^2}\, t'}.
\end{align}
Observing that $K^+_{1}<-\frac{1}{2 |b| \rho} $ when $t'> \frac{1} {4\rho^2 |b| \sqrt{1-b^2}}$, we know that the asymptotic soliton $u^+_{1}$ lies in the region between the direction $l_1: t'=-\frac{1}{2 |b| \rho} x $ and positive $t'$-axis, as seen in Fig.~\ref{Fig4}.

\item[(ii)]
If  $ W(x,t) =t'  e^{2\kappa\xi_1} = O(1)$, we take the limit of $u_{\rm{I\!I}}$ when $\xi_1 \ra -\infty $, yielding
\begin{subequations}
\begin{align}
&u_{\rm{I\!I}}\ra u^-_{1}=\rho  e^{2\, i \rho^2 t + i \phi}\bigg\{1 + \dfrac{4 b \sqrt{1-b^2} \omega \big[\gamma_1+4\rho^2b(1-b^2)^{\!\frac{3}{2}} t\,  e^{2\kappa(\xi_1+s_1)} \big]}{\gamma_1+8  i \rho^2 b^2(1-b^2)^{\!\frac{3}{2}}  e^{- i \theta} t\,  e^{2\kappa(\xi_1+s_1)}}\bigg\}, \label{2finala}\\
&|u_{\rm{I\!I}}|\ra|u^-_{1}|=\rho \bigg[1+\dfrac{2 \sqrt{1-b^2} \sin(\varphi_1-2\kappa s_{1I})}
{\sgn(b) \cosh(2\kappa\xi_1+ 2\kappa s_{1R}+\ln\Xi_1^-)- \sin(\theta+\varphi_1-2\kappa s_{1I})}\bigg]^{\frac{1}{2}},  \label{2finalb}
 \end{align}
 \end{subequations}
with  $\Xi_1^-=-8 \rho^2 b^2 (1-b^2)^{\!\frac{3}{2}}t'/r_1$. When condition~\eref{conda} is satisfied, $u^-_{1}$ is also nonsingular and it can represent an MAD soliton for $\sgn(b)\sin(\varphi_1-2\kappa s_{1I})>0$ or an MD soliton for $\sgn(b)\sin(\varphi_1-2\kappa s_{1I})<0$.
Since the asymptotic expression~\eref{2finala} is obtained by orderly letting $ \bar\xi_1, \xi_1\ra -\infty$, one immediately has
$\xi_1+\bar\xi_1 =4 \rho |b| t'\ra -\infty$. That is, $u^-_{1}$ appears as an asymptotic state of solution~\eref{3solution} as $t'\ra -\infty$. Via the extreme value analysis, the soliton center trajectory of $u^{-}_{1}$ can be determined as
\begin{align}
\mathcal{C}^-_{1}:\,\,t'  e^{\kappa\xi_1} =-\frac{r_1 e^{-2\kappa s_{1R}}} {8\rho^2 b^2 (1-b^2)^{\!\frac{3}{2}} }\quad (t' < 0), \label{Traj2}
\end{align}
and its slope  is given by
\begin{align}
K^-_{1}=-\frac{1}{2 |b|\rho }+\frac{1}{2 |b|\rho + 8\rho^3 b^2 \sqrt{1-b^2}\, t'},
\end{align}
which implies that $K^-_{1}<-\frac{1}{2 |b|\rho } $ when $t'<-\frac{1}{4 \rho^2 |b| \sqrt{1-b^2}}$. Therefore, the asymptotic soliton $u^-_{1}$ is located in the region between the direction $l_1$ and negative  $t'$-axis, as seen in Fig.~\ref{Fig4}.

\item[(iii)]
If  $ W(x,t) =t'  e^{2\kappa\xi_1} = O(1)$, we take the limit of $u_{\rm{I}}$ when $\xi_1\ra -\infty $, yielding
\begin{subequations}
\begin{align}
&u_{\rm{I}}\ra u^+_{2}=\rho  e^{2\, i \rho^2 t + i \phi}\bigg[1-\dfrac{ 16\rho^2(1-b^2)^2 t \, e^{2\kappa(\xi_1 + s_1) } }{8\rho^2 (1-b^2)^{\frac{3}{2}}  e^{-i \theta} t \, e^{2\kappa (\xi_1+s_1) } - i  b^2 \gamma_1}\bigg],\label{3finala}\\
& |u_{\rm{I}}|\ra |u^+_{2}|=\rho\bigg[1-\dfrac{2 \sqrt{1-b^2} \sin(\varphi_1-2\kappa s_{1I})}
{\sgn(b) \cosh(2\kappa\xi_1+2\kappa s_{1R} + \ln \Xi_2^+) + \sin(\theta+\varphi_1-2\kappa s_{1I})}\bigg]^{\frac{1}{2}},
\label{3finalb}
\end{align}
\end{subequations}
with $\Xi_2^+=8 \rho^2 (1-b^2)^{\!\frac{3}{2}}t'/ (b^2r_1)$. When $\varphi_1$, $b$ and $s_1$ satisfy
\begin{align}
\theta+\varphi_1-2\kappa s_{1I}\neq2k\pi-\frac{1}{2}\sgn(b)\pi \quad  (k \in \mathbb{Z}), \label{condb}
\end{align}
$u^+_{2}$ is nonsingular and it can represent an MAD soliton for $\sgn(b)\sin(\varphi_1-2\kappa s_{1I})<0$ or an MD soliton for $\sgn(b)\sin(\varphi_1-2\kappa s_{1I})>0$.
Note that the asymptotic expression~\eref{3finala} is obtained by orderly letting $\bar\xi_1\ra +\infty$ and $\xi_1\ra -\infty$,
which implies that $(\xi_1-\bar\xi_1)/2 = x \ra-\infty$.
Meanwhile, calculation of the extreme value of $|u^{+}_{2}|^2$ shows that the soliton center trajectory is
\begin{align}
\mathcal{C}^+_{2}:\,\, t'  e^{\kappa\xi_1} =\frac{b^2r_1  e^{-2\kappa s_{1R}}} {8\rho^2 (1-b^2)^{\!\frac{3}{2}} }\quad (t' >0), \label{Traj3}
\end{align}
where its slope is given by
\begin{align}
K^+_{2}=-\frac{1}{2 |b|\rho }+\frac{1}{2 |b|\rho + 8\rho^3 b^2 \sqrt{1-b^2}\, t'}.
\end{align}
Noticing that $-\frac{1}{2|b|\rho} < K^+_{2}<0 $ for $t' >0 $ and $ K^+_{2}< -\frac{1}{2|b|\rho} $ when $t' < \frac{-1 }{4\rho^2 |b| \sqrt{1-b^2}} $, we know that  $u^+_{2}$ must appear as an asymptotic soliton of solution~\eref{3solution} as $t'\ra +\infty$, and it is located in the region between the direction $l_1$ and negative  $x$-axis, as seen in Fig.~\ref{Fig4}.

\item[(iv)]
If  $ W(x,t) =t'  e^{-2\kappa\xi_1} = O(1)$,  we take the limit of $u_{\rm{I\!I}}$ when $\xi_1\ra +\infty $, yielding
\begin{subequations}
\begin{align}
&u_{\rm{I\!I}}\ra u^-_{2}=\rho  e^{2\, i \rho^2 t + i \phi}\bigg[1-\dfrac{ 16\rho^2(1-b^2)^2\gamma_1 t  e^{-2\kappa(\xi_1+s_1)} }{8 \rho^2 (1-b^2)^{\!\frac{3}{2}} \gamma_1  e^{ i \theta} t  e^{-2\kappa(\xi_1+s_1)} - i b^2  }\bigg], \label{4finala}\\
&|u_{\rm{I\!I}}|\ra|u^-_{2}|=\rho\bigg[1-\dfrac{2 \sqrt{1-b^2} \sin(\varphi_1-2\kappa s_{1I})}{\sgn(b)\cosh(2 \kappa\xi_1 + 2 \kappa s_{1R}-\ln\Xi_2^-)+ \sin(\theta+\varphi_1-2\kappa s_{1I})}\bigg]^{\frac{1}{2}},
 \label{4finalb}
\end{align}
\end{subequations}
with $\Xi_2^-=-8 \rho^2 (1-b^2)^{\!\frac{3}{2}} r_1 t'/ b^2$. When condition~\eref{condb} is  satisfied, $u^-_{2}$ is also nonsingular and it can represent an MAD soliton for $\sgn(b)\sin(\varphi_1-2\kappa s_{1I})<0$ or an MD soliton for $\sgn(b)\sin(\varphi_1-2\kappa s_{1I})>0$.
Since the asymptotic expression~\eref{4finala} is obtained by orderly letting $\bar\xi_1\ra -\infty$ and $\xi_1\ra +\infty$, one immediately have
$(\xi_1-\bar\xi_1)/2 = x \ra+\infty$.
Via the extreme value analysis, the soliton center trajectory of $u^{-}_{2}$ can be determined as
\begin{align}
\mathcal{C}^-_{2}:\,\,t'  e^{-\kappa\xi_1} =-\frac{b^2 e^{2\kappa s_{1R}}}{8\rho^2 (1-b^2)^{\!\frac{3}{2}} r_1}\quad (t' < 0), \label{Traj4}
\end{align}
and its slope is given by
\begin{align}
K^-_{2}=-\frac{1}{2 |b|\rho }+\frac{1}{2 |b|\rho -8\rho^3 b^2 \sqrt{1-b^2}\, t'},
\end{align}
which implies that $-\frac{1}{2\rho|b|}< K^-_{2} <0 $  for $t'<0 $ and $K^-_{2} < -\frac{1}{2\rho|b|} $  when $t' > \frac{1 }{4\rho^2 |b| \sqrt{1-b^2}} $. Therefore, $u^-_{2}$ must appear as an asymptotic soliton of solution~\eref{3solution} as $t'\ra -\infty$, and it is located  in the region between the direction $l_1$ and positive $x$-axis, as seen in Fig.~\ref{Fig4}.
\end{enumerate}

On the other hand, by precedently taking $\xi_1\ra\pm\infty$ for solution~\eref{3solution},  we have another two \emph{intermediate} asymptotic expressions:
\begin{align}
 u\ra u_{\rm{I\!I\!I}}= \,& \rho  e^{2\, i \rho^2 t +  i \phi} \notag \\
& \hspace{-5mm} \times\! \bigg\{1+ \frac{4 b \sqrt{1-b^2}\,\big[\omega^* e^{2\kappa(\bar\xi_1 +s_1^*)} +  i  \gamma^*_1  (2 i  b\sqrt{1-b^2} - \sqrt{1-b^2}\,\bar\zeta-1)\big]} {e^{2\kappa(\bar\xi_1 + s_1^*)}+2 i  b \sqrt{1-b^2}\, \gamma^*_1 ( e^{-2 i \theta}+ \bar\zeta e^{- i \theta}) +  i  b^4\omega {\gamma^*_1}^2  e^{-2\kappa(\bar\xi_1 +s_1^*)}}\bigg\}\,\,\, (\xi_1\ra \infty),  \label{5formsa} \\
 u\ra u_{\rm{I\!V}}=\, &\rho  e^{2\, i \rho^2 t + i \phi}  \notag \\
& \hspace{-5mm} \times \!\bigg\{1 + \frac{4b\sqrt{1-b^2}\,\big[\gamma^*_1   e^{2\kappa (\bar\xi_1 +s_1^*)}( \sqrt{1-b^2}\,\bar\zeta - 2 i  b \sqrt{1-b^2} - 1)+ i \omega {\gamma^*_1}^2 \big]}
{\omega^* b^4 e^{4\kappa(\bar\xi_1+ s_1^*)}+2 b \sqrt{1-b^2}\,\gamma^*_1   e^{2\kappa(\bar\xi_1+ s_1^*)}( e^{2 i \theta} - \bar\zeta e^{ i \theta})+ i  {\gamma^*_1}^2 }\bigg\}\,\,\, (\xi_1\ra -\infty), \label{6formsb}
\end{align}
with $\bar\zeta =   2 \rho[b^2(\bar\xi_1+s_1^*+s_2^*)-s_2^*] -4 \rho^2 |b|(1-b^2)t' $.
With an asymptotic analysis of Eqs.~\eref{5formsa} and~\eref{6formsb} like the above treatment on Eqs.~\eref{3formsa} and~\eref{3formsb}, the other four mixed asymptotic solitons of solution~\eref{3solution} can be obtained as follows:
\begin{enumerate}
\item[(i)]
Taking the limit of $u_{\rm{I\!I\!I}}$ when $\bar\xi_1\ra+\infty$,  the asymptotic expression along the curve $ W(x,t) =t' e^{-2\kappa\bar\xi_1}= O(1)$ is given by
\begin{subequations}
\begin{align}
 &u_{\rm{I\!I\!I}}\ra u^+_{3}=\rho  e^{2\, i \rho^2 t + i \phi}\bigg\{1+ \dfrac{4 b \sqrt{1-b^2}\,\big[ i \omega^{*} - 4 \rho^2 b(1-b^2)^{\!\frac{3}{2}} \gamma^*_1 t \, e^{-2\kappa(\bar\xi_1+s^*_1) }\big] }{ i + 8 \rho^2 b^2 (1-b^2)^{\!\frac{3}{2}} \gamma^*_1 e^{- i \theta} t \, e^{-2\kappa(\bar\xi_1+s^*_1) } }\bigg\},\label{5finala}\\
&|u_{\rm{I\!I\!I}}|\ra|u^+_{3}|=\rho\bigg[1+\dfrac{2\sqrt{1-b^2} \sin(2 \theta+\varphi_1-2\kappa s_{1I})}{\sgn(b) \cosh(2\kappa\bar\xi_1+ 2\kappa s_{1R}-\ln\Xi_1^+)-\sin(\theta+\varphi_1-2\kappa s_{1I})}\bigg]^{\frac{1}{2}},   \label{5finalb}
\end{align}
\end{subequations}
where the superscript ``$+$'' means $ t'\ra+\infty$.

\item[(ii)]
Taking the limit of $u_{\rm{I\!V}}$ when $\bar\xi_1\ra-\infty$, the asymptotic expression along the curve $ W(x,t) =t' e^{2\kappa\bar\xi_1}= O(1)$ is given by
\begin{subequations}
\begin{align}
&u_{\rm{I\!V}}\ra u^-_{3}=\rho  e^{2\, i \rho^2 t + i \phi}\bigg\{1+ \dfrac{4 b \sqrt{1-b^2} \big[i \omega \gamma^*_1 - 4 \rho^2 b(1-b^2)^{\!\frac{3}{2}} t  e^{2\kappa(\bar\xi_1+s^*_1)} \big]}{ i \gamma^*_1+8 \rho^2 b^2(1-b^2)^{\!\frac{3}{2}}  e^{ i \theta} t e^{2\kappa(\bar\xi_1 + s^*_1)} }\bigg\},\label{6finala}\\
& |u_{\rm{I\!V}}|\ra|u^-_{3}|=\rho\bigg[1+\dfrac{2\sqrt{1-b^2} \sin(2 \theta+\varphi_1-2\kappa s_{1I})}{\sgn(b) \cosh(2\kappa\bar\xi_1+ 2\kappa s_{1R}+\ln\Xi_1^-) -\sin(\theta+\varphi_1-2\kappa s_{1I})}\bigg]^{\frac{1}{2}}, \label{6finalb}
\end{align}
\end{subequations}
where the superscript ``$-$'' means $ t'\ra-\infty$.

\item[(iii)]
Taking the limit of $u_{\rm{I\!I\!I}}$ when $\bar\xi_1\ra-\infty$, the asymptotic expression along the curve  $ W(x,t) =t' e^{2\kappa\bar\xi_1}= O(1)$ is given by
\begin{subequations}
\begin{align}
& u_{\rm{I\!I\!I}}\ra u^+_{4}=\rho  e^{2\, i \rho^2 t + i \phi}\bigg[1-\dfrac{ 16 \rho^2 (1-b^2)^2 t e^{2\kappa(\bar\xi_1+s^*_1)}}{8 \rho^2 (1-b^2)^{\!\frac{3}{2}}  e^{- i \theta} t  e^{2\kappa(\bar\xi_1+s^*_1)}- b^2 \omega \gamma^*_1}\bigg],\label{7finala}\\
& |u_{\rm{I\!I\!I}}|\ra|u^+_{4}|=\rho\bigg[1-\dfrac{2\sqrt{1-b^2} \sin(2 \theta+\varphi_1-2\kappa s_{1I}) }{\sgn(b) \cosh(2\kappa\bar\xi_1+ 2\kappa s_{1R}+\ln\Xi_2^+)+\sin(\theta+\varphi_1-2\kappa s_{1I})}\bigg]^{\frac{1}{2}}, \label{7finalc}
\end{align}
\end{subequations}
where the superscript ``$+$'' means $ t'\ra +\infty$.

\item[(iv)]
Taking the limit of $u_{\rm{I\!V}}$ when $\bar\xi_1\ra+\infty$, the asymptotic expression along the curve  $ W(x,t) =t' e^{-2\kappa\bar\xi_1}= O(1)$ is given by
\begin{subequations}
\begin{align}
&u_{\rm{I\!V}}\ra u^-_{4}=\rho  e^{2\, i \rho^2 t + i \phi}\bigg[1-\dfrac{ 16 \rho^2 (1-b^2)^2 \gamma^*_1 t e^{-2\kappa(\bar\xi_1+s^*_1) }}{8 \rho^2 (1-b^2)^{\!\frac{3}{2}} \gamma^*_1  e^{ i \theta} t  e^{-2\kappa(\bar\xi_1+s^*_1) }+ b^2 \omega^* }\bigg],\label{8finala} \\
&|u_{\rm{I\!V}}|\ra|u^-_{4}|=\rho\bigg[1-\dfrac{2\sqrt{1-b^2} \sin(2 \theta+\varphi_1-2\kappa s_{1I})}{\sgn(b) \cosh(2\kappa\bar\xi_1+ 2\kappa s_{1R}-\ln\Xi_2^-)+\sin(\theta+\varphi_1-2\kappa s_{1I})}\bigg]^{\frac{1}{2}},\label{8finalb}
\end{align}
\end{subequations}
where the superscript ``$-$'' means $ t'\ra-\infty$.
\end{enumerate}

As seen from Eqs.~\eref{5finala}--\eref{8finalb}, we know that $u^{\pm}_{3}$ is nonsingular with condition~\eref{conda} and it can describe an MAD soliton for $\sgn(b)\sin(2 \theta+\varphi_1-2\kappa s_{1I})>0$ or an MD soliton for $\sgn(b)\sin(2 \theta+\varphi_1-2\kappa s_{1I})<0$; $u^{\pm}_{4}$ is nonsingular with condition~\eref{condb} and it can describe an MAD soliton  for $\sgn(b)\sin(2 \theta+\varphi_1-2\kappa s_{1I})<0$ or an MD soliton for $\sgn(b)\sin(2 \theta+\varphi_1-2\kappa s_{1I})>0$.
Meanwhile,  the center trajectories and their slopes of the asymptotic solitons $u^\pm_{3}$ and $u^\pm_{4}$ can be given as follows:
\begin{subequations}
\begin{align}
&\mathcal{C}^+_{3}:\,\,t'  e^{-2\kappa\bar\xi_1} =\frac{ e^{2\kappa s_{1R}}}{8\rho^2 b^2(1-b^2)^{\!\frac{3}{2}} r_1 },\quad  K^+_{3}=\frac{1}{2\rho|b|}-\frac{1}{2\rho|b|-8\rho^3b^2\sqrt{1-b^2}\,t'},   \label{TrSl34a} \\
&\mathcal{C}^-_{3}:\,\,t'  e^{2\kappa\bar\xi_1} =-\frac{r_1 e^{-2\kappa s_{1R}}} {8\rho^2 b^2 (1-b^2)^{\!\frac{3}{2}} },\quad \,\,\, K^-_{3}=\frac{1} {2\rho|b|}- \frac{1}{2\rho|b|+ 8\rho^3 b^2\sqrt{1-b^2}\, t'},  \label{TrSl34b} \\
&\mathcal{C}^+_{4}:\,\,t'  e^{2\kappa\bar\xi_1} =\frac{b^2 r_1  e^{-2\kappa s_{1R}}} {8\rho^2 (1-b^2)^{\!\frac{3}{2}} },\quad \quad \,\,\,\,\,\, K^+_{4}=\frac{1}{2\rho|b|}-\frac{1}{2\rho|b|+8\rho^3 b^2 \sqrt{1-b^2}\, t'}, \label{TrSl34c} \\
&\mathcal{C}^-_{4}:\,\,t'  e^{-2\kappa\bar\xi_1} =-\frac{b^2  e^{2\kappa s_{1R}}}{8\rho^2 (1-b^2)^{\!\frac{3}{2}} r_1},\quad  K^-_{4}=\frac{1}{2\rho|b|}-\frac{1}{2\rho|b|-8\rho^3 b^2 \sqrt{1-b^2}\,t'}. \label{TrSl34d}
\end{align}
\end{subequations}
It can be readily obtained that the slope $K^+_{3}>\frac{1}{2 |b|\rho } $ when $t'>\frac{1} {4\rho^2 |b| \sqrt{1-b^2}}$, $K^-_{3}>\frac{1}{2 |b|\rho } $ when $t'< -\frac{1} {4\rho^2 |b| \sqrt{1-b^2}}$, $0<K^+_{4}<\frac{1}{2 |b|\rho } $ when  $t'>0$, and $0<K^-_{4}<\frac{1}{2 |b|\rho } $ when $t'<0$. That is, the asymptotic soliton $u^+_{3}$ is situated between the direction $l_2:  t'=\frac{1}{2 |b| \rho} x $ and positive $t'$-axis,  $u^-_{3}$ is between the direction $l_2$ and negative $t'$-axis,  $u^+_{4}$ is between the direction $l_2$ and positive $x$-axis, and $u^-_{4}$ is between the direction $l_2$ and negative $x$-axis. The distribution of  four asymptotic soliton pairs in the $xt'$ plane can be found in Fig.~\ref{Fig4}.

\begin{figure}[ht]
\centering
\subfigure[]{\label{Fig5a}
\includegraphics[width=5.5in]{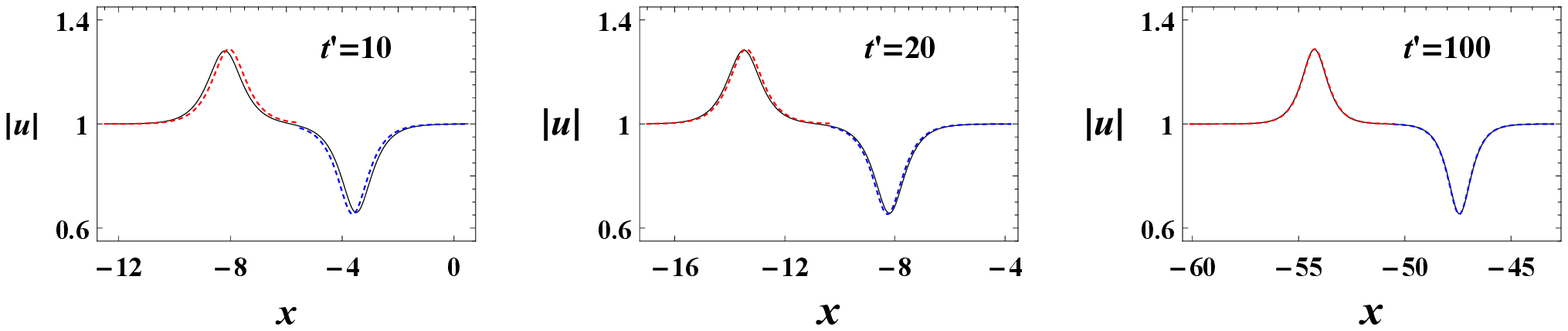}}
\subfigure[]{\label{Fig6a}
\includegraphics[width=5.5in]{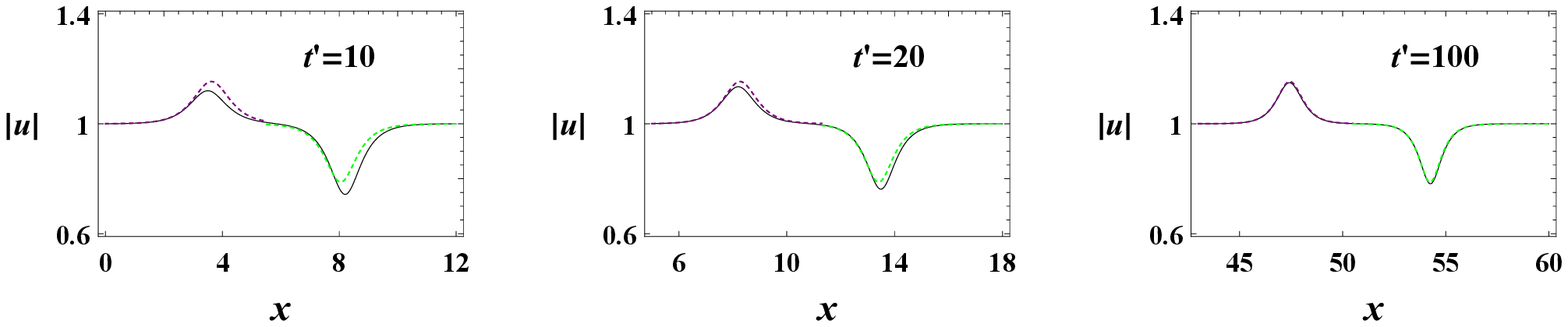}}
\subfigure[]{\label{Fig5b}
\includegraphics[width=5.5in]{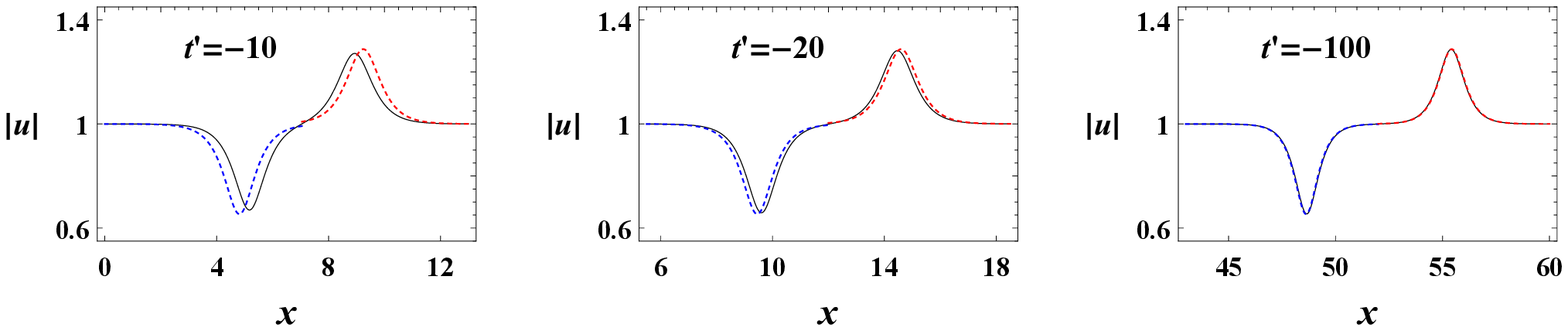}}
\subfigure[]{\label{Fig6b}
\includegraphics[width=5.5in]{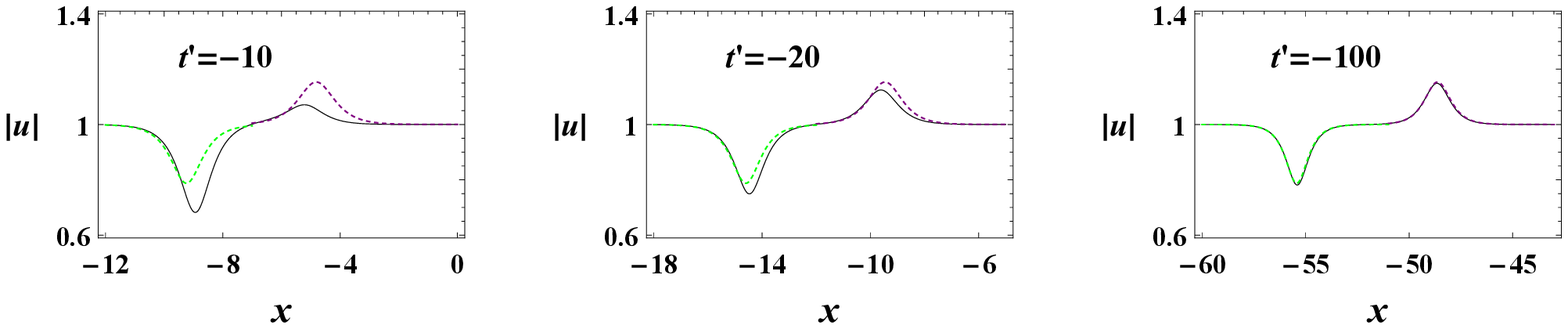}}
\caption{  (a) Comparison of the asymptotic solitons $u^+_1$ (blue dashed) and $u^+_{2}$ (red dashed) with the exact solution~\eref{3solution} (black solid). (b) Comparison of the asymptotic solitons $u^+_3$ (purple dashed) and $u^+_4$ (green dashed) with the exact solution~\eref{3solution} (black solid).
(c) Comparison of the asymptotic solitons $u^-_1$ (blue dashed) and $u^-_{2}$ (red dashed) with the exact solution~\eref{3solution} (black solid). (d) Comparison of the asymptotic solitons $u^-_3$ (purple dashed) and $u^-_4$ (green dashed) with the exact solution~\eref{3solution} (black solid). The relevant parameters are selected as $\rho=1$, $b=\frac{1}{4}$, $\phi=0$, $s_1=0$, $s_2=2$ and $\gamma_1=3- i $. \label{Fig5}}
\end{figure}

The above asymptotic analysis of solution~\eref{3solution} is apparently more complicated than that of solution~\eref{2solution}, and unconventionally the center trajectories of eight asymptotic solitons $u^\pm_i$ ($1\leq i\leq 4$)  are all along some curved lines in the $xt'$ plane. Naturally, one may ask how well those asymptotic expressions approximate the exact solution when $|t'| \gg 1$.
In Fig.~\ref{Fig5}, we compare the asymptotic solitons  $u^\pm_i$  ($1\leq i\leq 4$)  with the exact solution~\eref{3solution} at different values of $t'$. The graphical comparison shows that the asymptotic expressions give a good approximation to solution~\eref{3solution} for large values of $t'$.

\subsection{Properties of soliton interactions}

\begin{figure}[ht]
 \centering
{\includegraphics[width=2.8in]{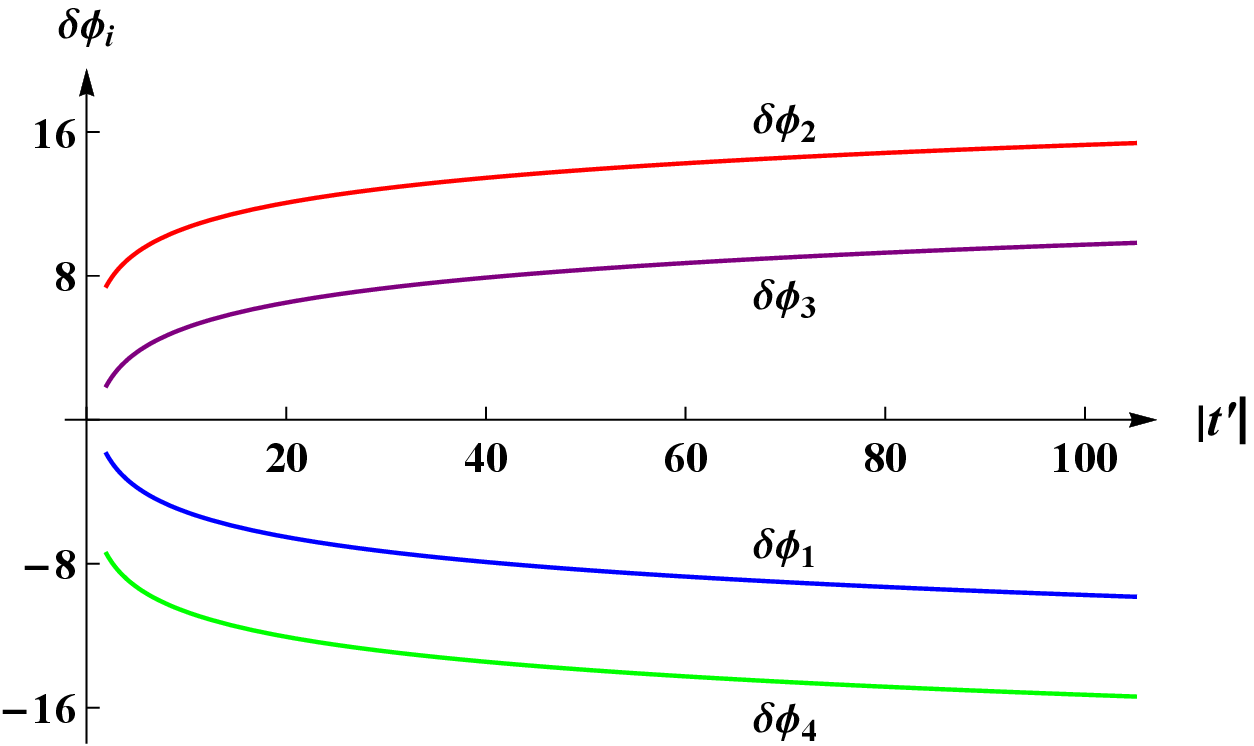}}\hfill
\caption{ The phase differences for the envelopes of $u_i^{+}$ and $u_i^{-}$ ($1\leq i \leq 4$) versus $t'$, where the parameters are selected as $\rho=1$, $b=\frac{1}{2}$ and $\phi=0$.  \label{Fig7} }
\end{figure}

Based on the obtained asymptotic expressions in Subsection~\ref{sec4.1}, we discuss the soliton interaction properties of solution~\eref{3solution} in the following aspects:
\begin{enumerate}
\item[(i)]
By calculating the absolute differences between $|u^\pm_i|_{\rm{max}}^2$ and $\rho^2$ for the MAD solitons  (or between  $|u^\pm_i|_{\rm{min}}^2$ and $\rho^2$ for the MD solitons), we get the amplitudes for $|u^\pm_i|^2$ ($1\leq i\leq 4$) as follows:
\begin{subequations}
\begin{align}
&A_1^{\pm}=\frac{2\rho^2 \sqrt{1-b^2} |\sin(\varphi_1-2\kappa s_{1I})|}{1-\sgn(b)\sin(\theta+\varphi_1-2\kappa s_{1I})}, \quad\quad\,\,\,  A_{2}^{\pm}=\frac{2\rho^2 \sqrt{1-b^2} |\sin(\varphi_1-2\kappa s_{1I})| }{
1+\sgn(b)\sin(\theta+\varphi_1-2\kappa s_{1I})},\label{6zf}\\
& A_{3}^{\pm}= \frac{2\rho^2\sqrt{1-b^2}|\sin(2 \theta+\varphi_1-2\kappa s_{1I})|}{1-\sgn(b)\sin(\theta+\varphi_1-2\kappa s_{1I})},\quad A_{4}^{\pm}=
\frac{2\rho^2\sqrt{1-b^2}|\sin(2 \theta+\varphi_1-2\kappa s_{1I})|} {1+
\sgn(b)\sin(\theta+\varphi_1-2\kappa s_{1I})},\label{8zf}
\end{align}\label{4forms}
\end{subequations}
which shows that each pair of asymptotic solitons $(u^+_i, u^-_i)$ have the same amplitudes.

\item[(ii)]

Since the center trajectories of all asymptotic solitons are  along some curved lines in the $xt'$ plane,   $u^\pm_i$ ($1\leq i\leq 4$) have the $t'$-dependent velocities, namely,
\begin{subequations}
\begin{align}
& v_{1}^{\pm}=-2 |b|\rho +\frac{1}{2\rho\sqrt{1-b^2}|t'|}, \quad v_{2}^{\pm}=-2 |b|\rho -\frac{1}{2\rho\sqrt{1-b^2}|t'|},\\
& v_{3}^{\pm}=\,2 |b|\rho -\frac{1}{2\rho\sqrt{1-b^2}|t'|}, \quad\,\,\,\, v_{4}^{\pm}=\,2 |b|\rho +\frac{1}{2\rho\sqrt{1-b^2}|t'|}.
\end{align}
\end{subequations}
The absolute differences $|v_{1}^{\pm}-v_{2}^{\pm}|$ and $|v_{3}^{\pm}-v_{4}^{\pm}|$ will increase when  $t'$ ranges from  $+\infty$ to $0$, which implies that the attraction between $u^\pm_1$ and $u^\pm_2$ (or between $u^\pm_3$ and $u^\pm_4$) gradually gets strengthened in the near-field region. But as $|t'|\ra +\infty$, $ v_{1,2}^{\pm} \ra -2 |b|\rho $ and $ v_{3,4}^{\pm} \ra 2 |b|\rho $, so that the asymptotic solitons $u^\pm_1$ and $u^\pm_2$ ($u^\pm_3$ and $u^\pm_4$) tend to be parallel to each other in the far-field region.

\item[(iii)]
There still exist the  phase shifts for the envelopes of $u_i^{+}$ and $u_i^{-}$ ($1\leq i\leq4$), and the phase differences can be given by
\begin{subequations}
\begin{align}
&\delta\phi_{1}=-\delta\phi_{3}=-\ln\Xi_1^+-\ln\Xi_1^- =-2 \ln[8\rho^2 b^2(1-b^2)^{\!\frac{3}{2}}|t'|], \label{delphi2a} \\[1mm]
&\delta\phi_{2}=-\delta\phi_{4}=\ln\Xi_2^{+} + \ln\Xi_2^- =2\ln[8\rho^2(1-b^2)^{\!\frac{3}{2}}|t'|/b^2]. \label{delphi2b}
\end{align}  \label{delphi2}
\end{subequations}
In contrast to the fixed phase shifts in Eq.~\eref{delphi1}, the absolute phase differences $|\delta\phi_{i}|$  in Eqs.~\eref{delphi2a} and~\eref{delphi2b}  grow as  $|t'|$ increases in the logarithmical manner (see Fig.~\ref{Fig7}).

\end{enumerate}

\begin{figure}[ht]
 \centering
\subfigure[]{\label{Fig8a}
\includegraphics[width=1.55in]{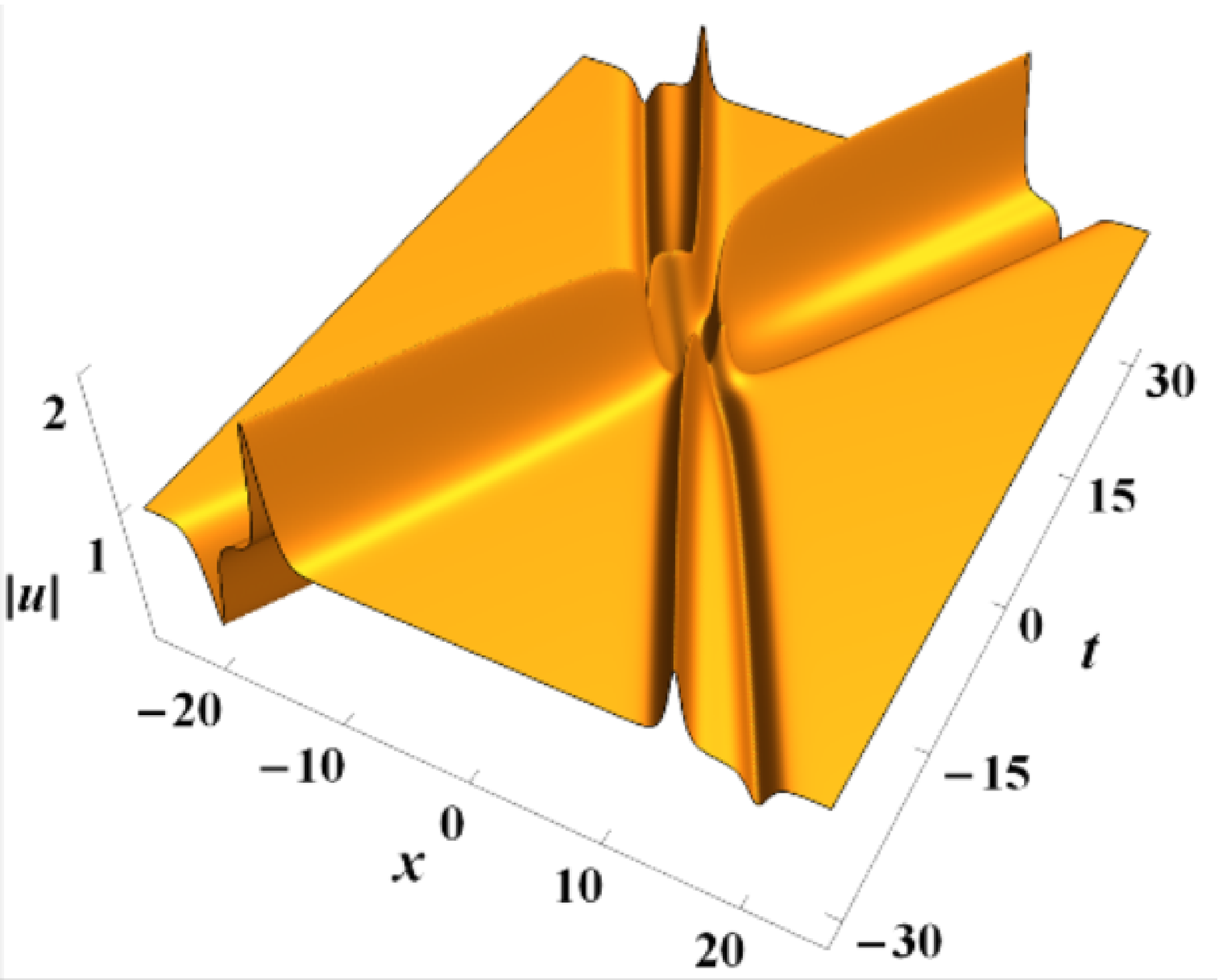}}\hfill
\subfigure[]{ \label{Fig8b}
\includegraphics[width=1.55in]{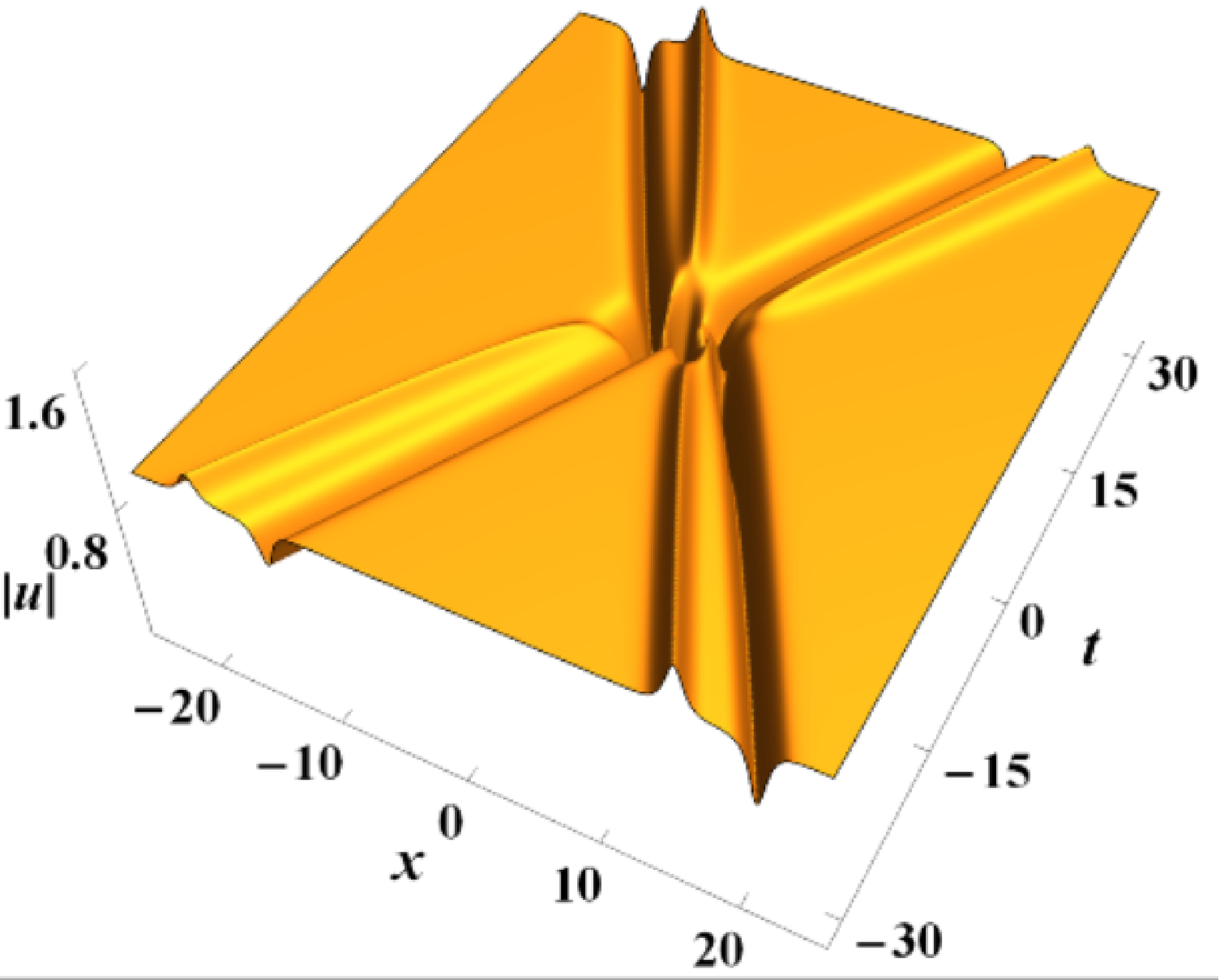}}\hfill
\subfigure[]{ \label{Fig8c}
\includegraphics[width=1.55in]{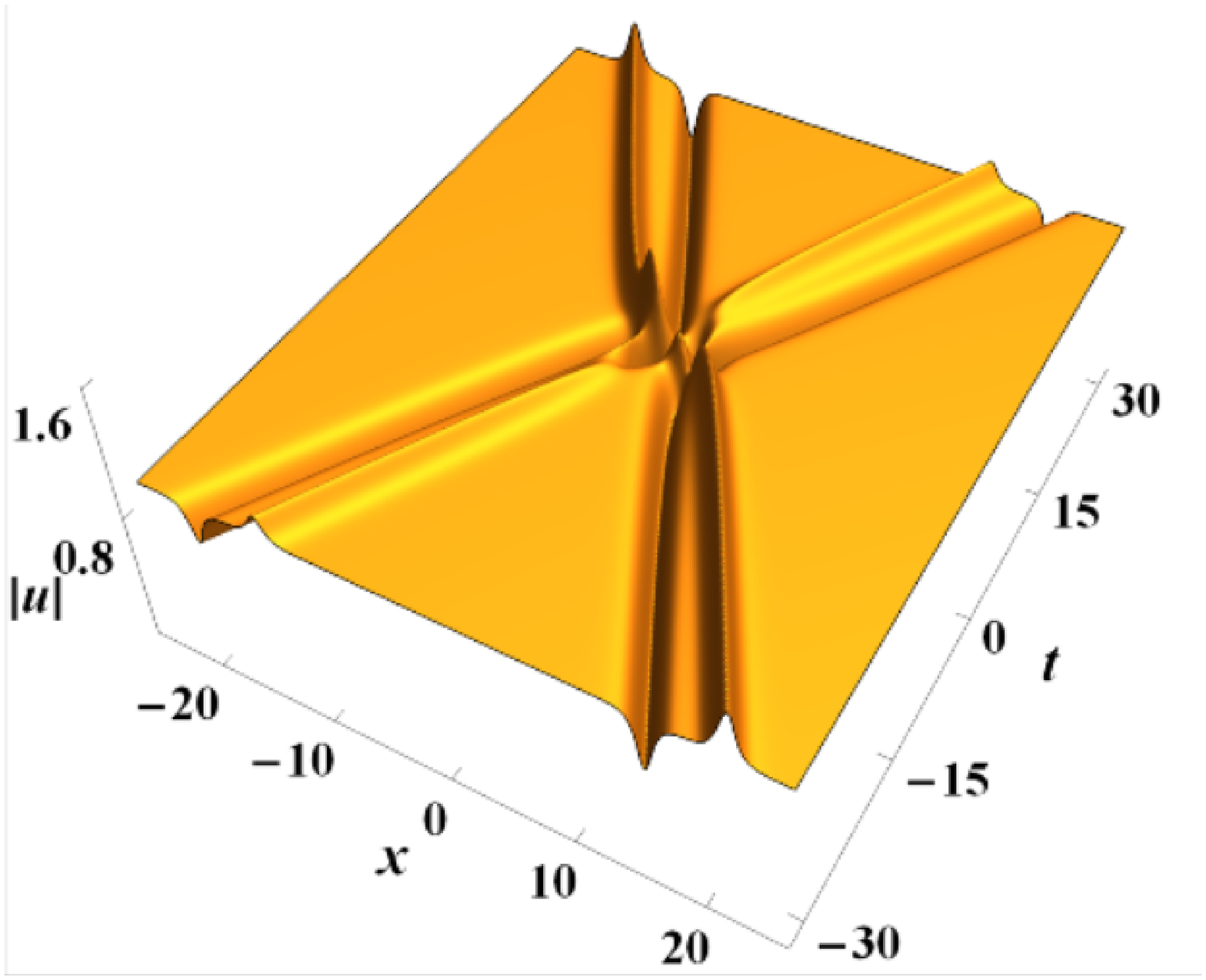}}\hfill
\subfigure[]{\label{Fig8d}
\includegraphics[width=1.55in]{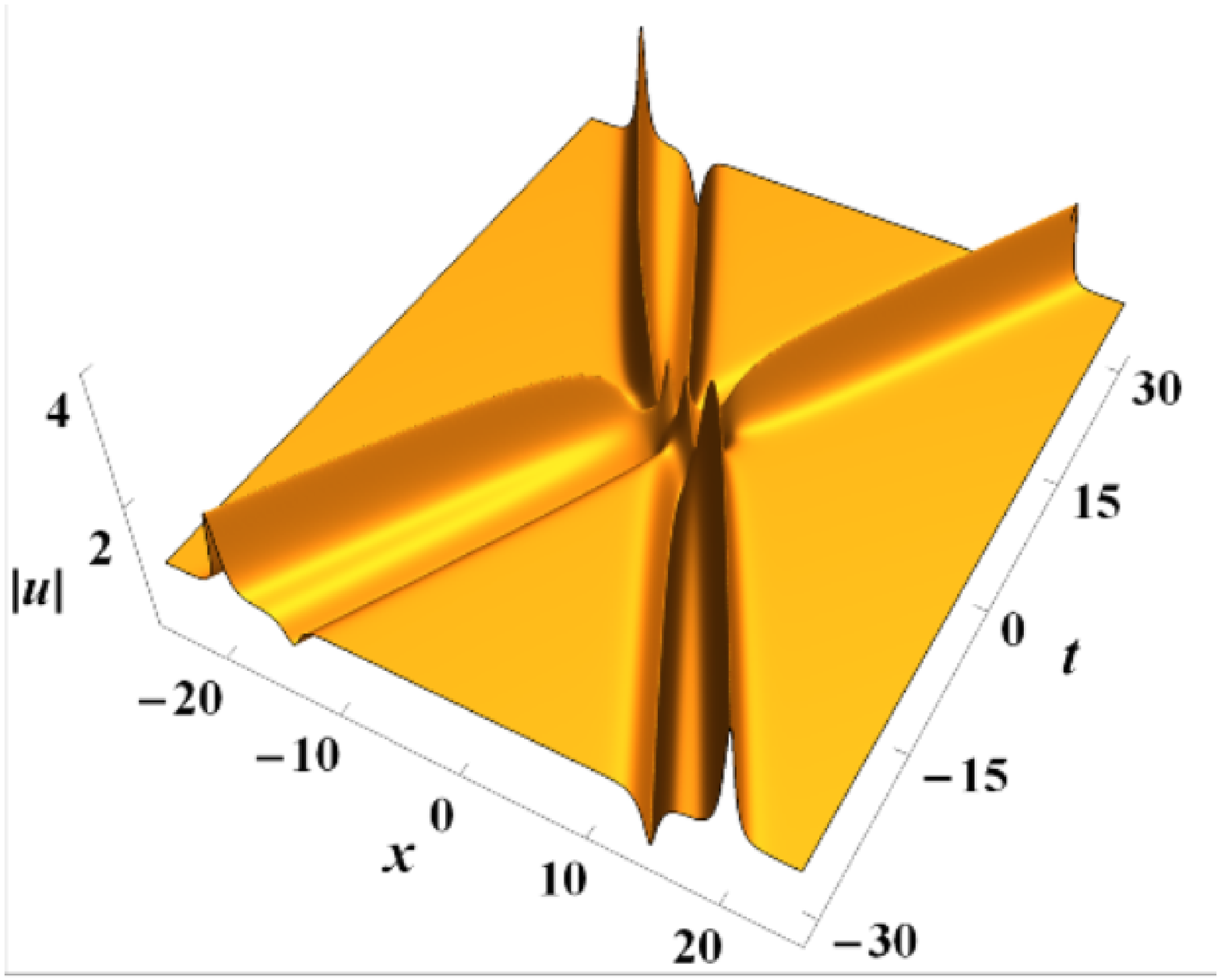}}\hfill
\caption{\small  Four types of non-degenerate four-soliton interactions via solution~\eref{3solution}:  (a) MAD-MD-MAD-MD soliton interaction with $\rho=1$, $b=\frac{1}{4}$, $\phi=0$, $s_1=0$, $s_2=0$ and $\gamma_1=3+ i $. (b)  MAD-MD-MD-MAD soliton interaction with $\rho=1$, $b=\frac{1}{4}$, $\phi=0$, $s_1=0$, $s_2=\frac{1}{2} i $ and $\gamma_1=-3+ i $. (c) MD-MAD-MAD-MD soliton interaction with $\rho=1$, $b=\frac{1}{4}$, $\phi=0$, $s_1=0$, $s_2=0$ and $\gamma_1=3- i $. (d) MD-MAD-MD-MAD soliton interaction with $\rho=1$, $b=\frac{1}{4}$, $\phi=0$, $s_1=0$, $s_2=0$ and $\gamma_1=\frac{1}{2}- i $. \label{Fig8} }
\end{figure}

\begin{figure}[ht]
 \centering
\subfigure[]{\label{Fig9a}
\includegraphics[width=1.55in]{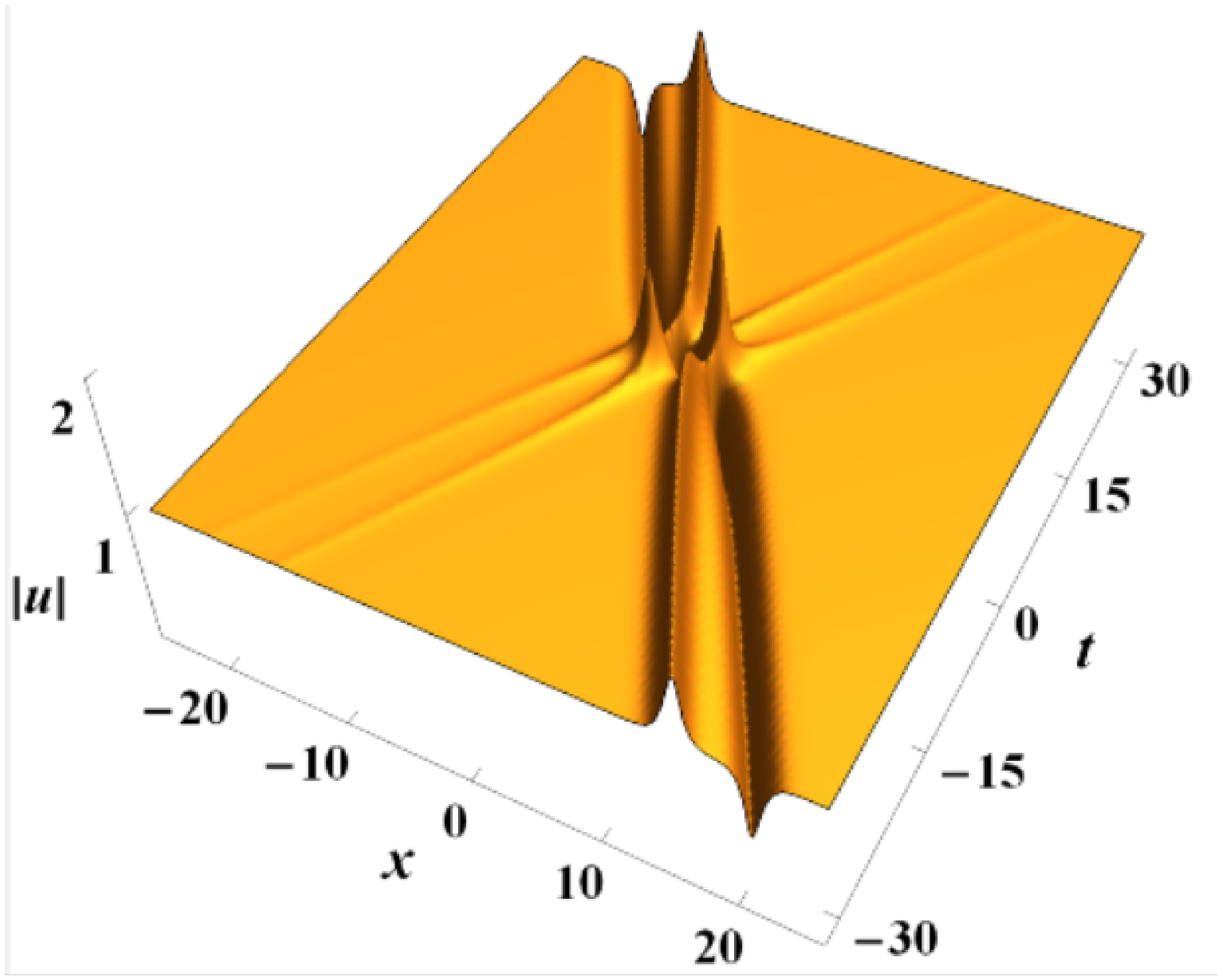}}\hfill
\subfigure[]{ \label{Fig9b}
\includegraphics[width=1.55in]{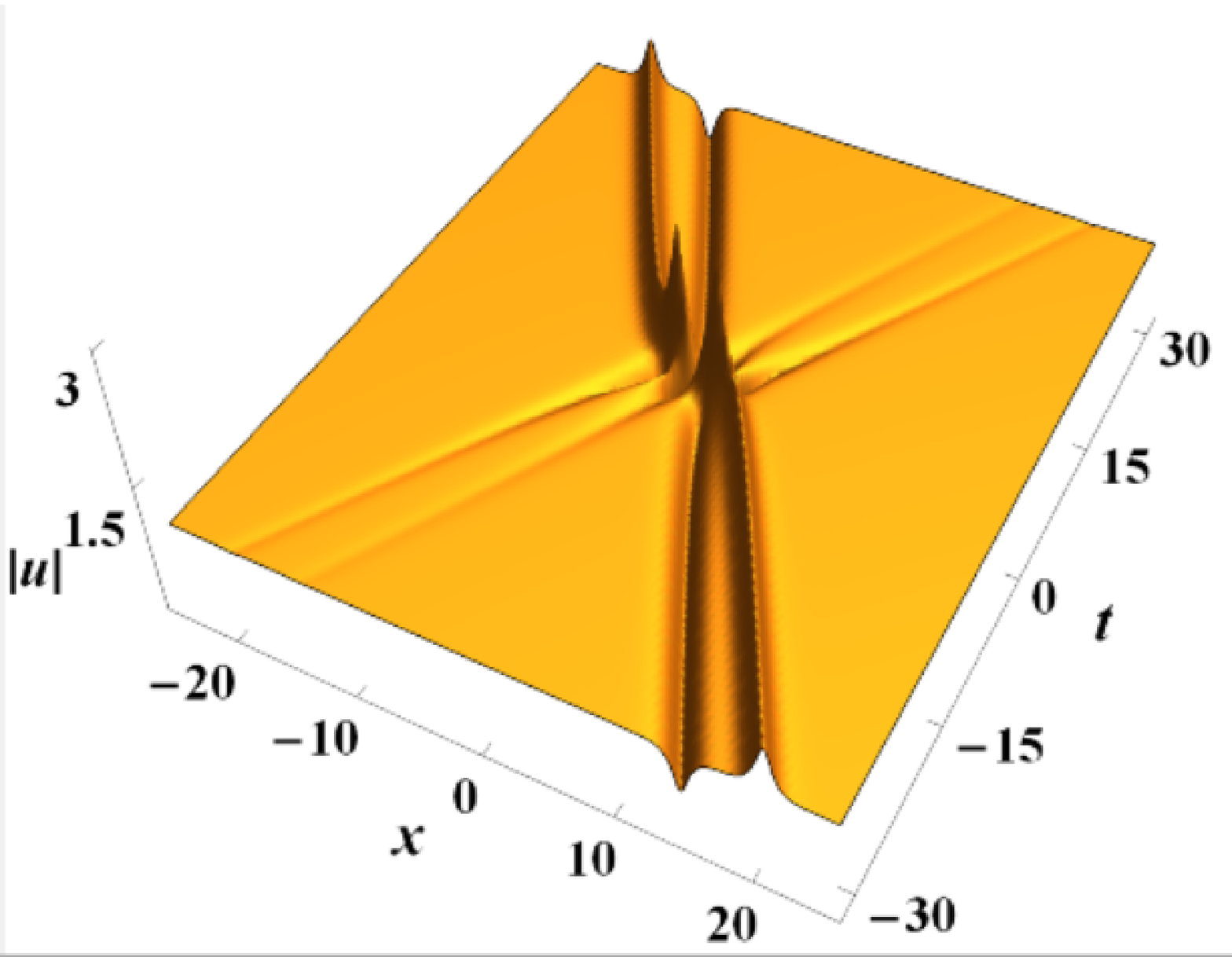}}\hfill
\subfigure[]{ \label{Fig9c}
\includegraphics[width=1.55in]{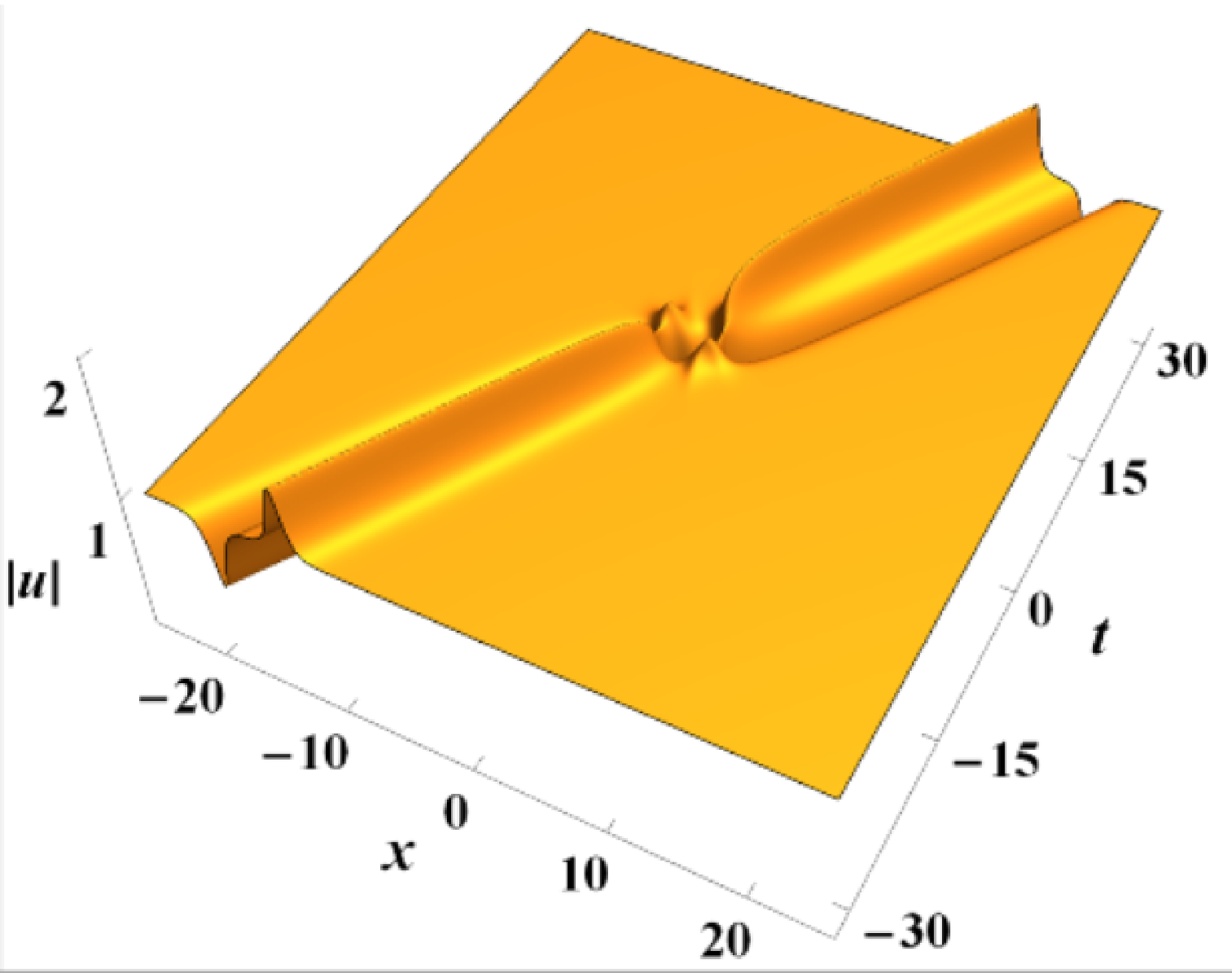}}\hfill
\subfigure[]{\label{Fig9d}
\includegraphics[width=1.55in]{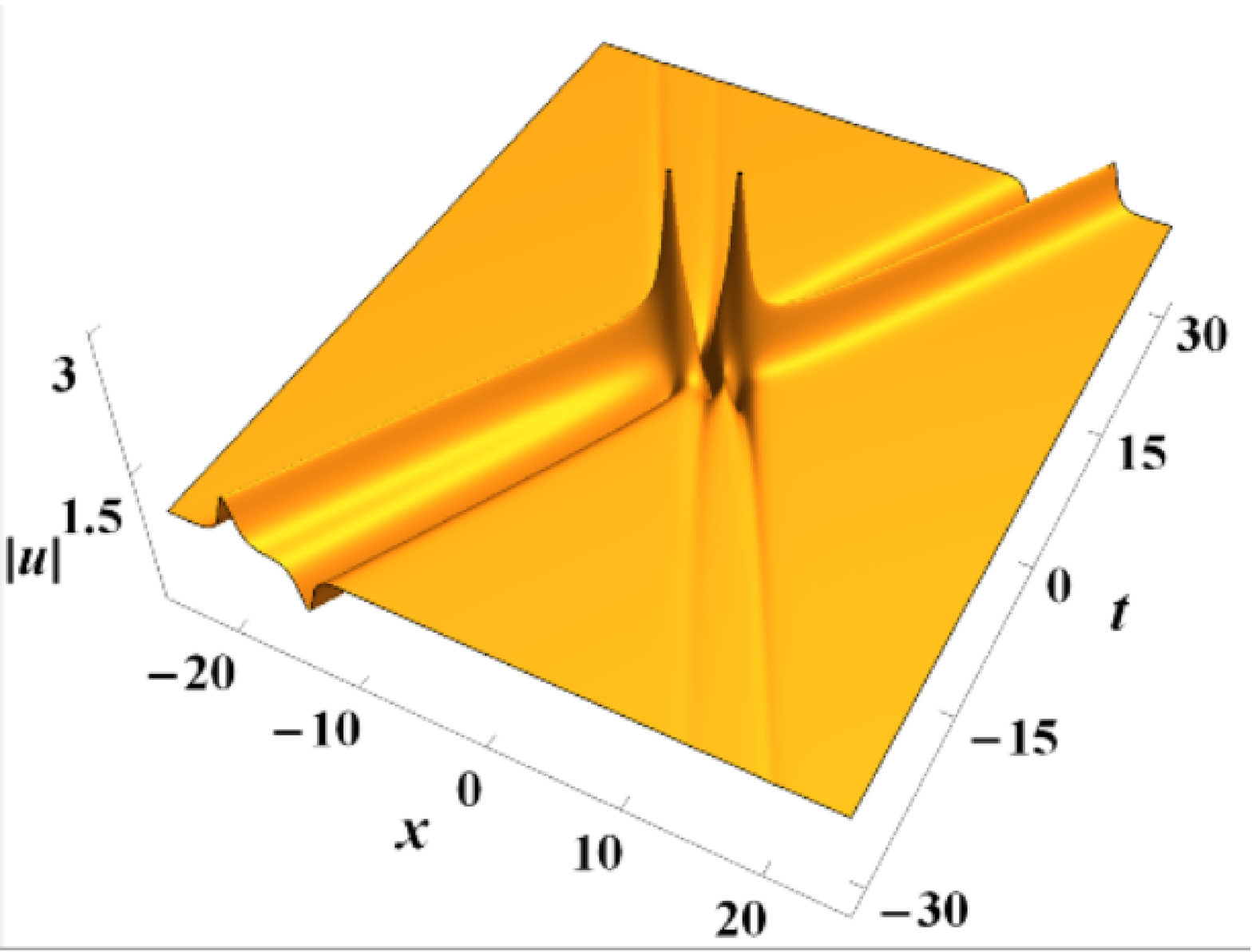}}\hfill
\caption{\small Four types of degenerate four-soliton interactions via
solution~\eref{3solution}:  (a) MAD-MD-V-V interaction with $\rho=1$, $b=\frac{1}{4}$, $\phi=0$, $s_1=0$, $s_2=\frac{1}{5} i $ and $\gamma_1=-1+\frac{\sqrt{15}}{7} i $. (b) MD-MAD-V-V interaction with $\rho=1$, $b=\frac{1}{4}$, $\phi=0$, $s_1=0$, $s_2=\frac{3}{25}i$ and $\gamma_1=1-\frac{\sqrt{15}}{7} i $. (c) V-V-MAD-MD  interaction with $\rho=1$, $b=\frac{1}{4}$, $\phi=0$, $s_1=0$, $s_2=0$ and $\gamma_1=1$. (d) V-V-MD-MAD interaction with $\rho=1$, $b=\frac{1}{4}$, $\phi=0$, $s_1=0$, $s_2=\frac{3}{25} i $ and $\gamma_1=-1$.  Here, ``V'' represents the vanishment of an asymptotic soliton as $|t|\ra \infty$.   \label{Fig9} }
\end{figure}

From the above analysis, we can regard the soliton interactions in solution~\eref{3solution} are still \emph{elastic} in the sense that the  shapes and amplitudes of all interacting solitons are retained upon interaction, and that $u^+_i$ has the same velocity at $t'$ as that of $u^-_i$ at $-t'$.  Although each asymptotic soliton profile could be either the dark or antidark type, solution~\eref{3solution} admits only four types of non-degenerate four-soliton interactions, as shown in  Figs.~\ref{Fig8a}--\ref{Fig8d}. Note that $u^\pm_1$ and $u^\pm_2$ always have the opposite soliton profiles, so do  $u^\pm_3$ and $u^\pm_4$ (see Table~\ref{Table3}). This is because the parametric conditions determining the soliton profiles of $u_i^\pm$ ($1\leq i\leq4$) are not independent of each other. Besides, the four-soliton interaction can only degenerate to a two-soliton interaction, that is, both $u^{\pm}_1$ and $u_2^{\pm}$ vanish as $t'\ra\pm \infty$ for $ \sin(\varphi_1-2\kappa s_{1I})=0$, or both $u^{\pm}_3$ and $u_4^{\pm}$ vanish as $t'\ra\pm \infty$ for $\sin(2 \theta+\varphi_1-2\kappa s_{1I})=0$. In Figs.~\ref{Fig9a}--\ref{Fig9d}, we depict all the four types of degenerate four-soliton interactions described by solution~\eref{3solution}.  Some small ripples can be observed in the near-field region, but they will eventually disappear as long as  $|t'|$ is large enough. Accordingly, the degenerate cases cannot be regarded as the conventional two-soliton interactions, neither.


\begin{table}[h]
\small
\caption{Types of mixed asymptotic solitons in solution~\eref{3solution} with
different parametric conditions. \label{Table3}} \vspace{0mm}
\begin{center}
\begin{tabular}{|c|c|c|c|c|}
\hline  \multicolumn{1}{|c|}{Parametric conditions} & { $u_1^{\pm}$}& { $u_2^{\pm}$}& { $u_3^{\pm}$}& { $u_4^{\pm}$}
\\ \hline
\tabincell{c}{$\sgn(b)\sin(\varphi_1-2\kappa s_{1I})>0$,  \\ $\sgn(b)\sin(2 \theta+\varphi_1-2\kappa s_{1I})>0$}   &  MAD soliton  &  MD soliton  &  MAD soliton  &  MD soliton
\\ \hline
\tabincell{c}{$\sgn(b)\sin(\varphi_1-2\kappa s_{1I})>0$,\\ $\sgn(b)\sin(2 \theta+\varphi_1-2\kappa s_{1I})<0$} &  MAD soliton  &  MD soliton  &  MD soliton  &  MAD soliton
\\ \hline
\tabincell{c}{$\sgn(b)\sin(\varphi_1-2\kappa s_{1I})<0$,  \\ $\sgn(b)\sin(2 \theta+\varphi_1-2\kappa s_{1I})>0$} &  MD soliton  &  MAD soliton  &  MAD soliton  &  MD soliton
\\ \hline
\tabincell{c}{$\sgn(b)\sin(\varphi_1-2\kappa s_{1I})<0$,  \\ $\sgn(b)\sin(2 \theta+\varphi_1-2\kappa s_{1I})<0$} &  MD soliton  &  MAD soliton  &  MD soliton  &  MAD soliton
\\ \hline
\tabincell{c}{$\sgn(b)\sin(\varphi_1-2\kappa s_{1I})>0$,  \\ $\sin(2 \theta+\varphi_1-2\kappa s_{1I})=0$}   &  MAD soliton  &  MD soliton  &  Vanish   &Vanish
\\ \hline
\tabincell{c}{$\sgn(b)\sin(\varphi_1-2\kappa s_{1I})<0$,  \\ $\sin(2 \theta+\varphi_1-2\kappa s_{1I})=0$} &  MD soliton  &  MAD soliton  &  Vanish   &Vanish
\\ \hline
\tabincell{c}{$\sin(\varphi_1-2\kappa s_{1I})=0$, \\$ \sgn(b)\sin(2 \theta+\varphi_1-2\kappa s_{1I})>0$}  &  Vanish   &Vanish   &  MAD soliton  & MD soliton
\\ \hline
\tabincell{c}{$\sin(\varphi_1-2\kappa s_{1I})=0$, \\$\sgn(b) \sin(2 \theta+\varphi_1-2\kappa s_{1I})<0$}  &  Vanish   &Vanish   &  MD soliton  & MAD soliton
 \\ \hline
\end{tabular}
\end{center}
\end{table}

\section{Conclusions and discussions} \label{Sec5}

It has been shown in Refs.~\cite{LiXu,LiXu1} that the defocusing nonlocal NLS equation  admits both the exponential and rational soliton solutions on the cw background $u_{\rm{cw}}=\rho\,e^{i\,(2 \rho^2 t +\phi)}$. In this paper, by using the twice-iterated DT and starting from the same seed $u_{\rm{cw}}$,  we have constructed two new types of exponential-and-rational mixed soliton solutions for Eq.~\eref{NNLS} with $\varepsilon=-1$. Via the asymptotic analysis method, we have revealed that there are two exponential solitons and two rational ones in the first type of solution, and four mixed  solitons in the second type of solution. The two types of solutions can exhibit a variety of elastic four-soliton interactions since each asymptotic soliton could be either the dark or antidark type. Also, we have discussed the degenerate cases when the four-soliton interaction reduces to a three-soliton  or two-soliton interaction. For such two types of mixed soliton solutions,  we have given the parametric conditions associated with all possible types of soliton interactions in Tables~\ref{Table1}--\ref{Table3}. Specially, we have revealed that the asymptotic solitons in the second type of solution have the $t$-dependent velocities and their phase shifts  before and after interaction also grow with $|t|$ in the logarithmical manner, which is in sharp contrast with that in the local NLS equation.  Finally, we would like to discuss the following issues:

\begin{enumerate}

\item[(i)]
It is a challenging work to find the sufficient and necessary nonsingular conditions for  the soliton solutions of an integrable nonlocal equation~\cite{LiXu1,LiXu3}. Although all the asymptotic solitons of solution~\eref{2solution} (or solution~\eref{3solution}) are globally nonsingular if and only if conditions~\eref{cond1} and~\eref{cond2} (or conditions~\eref{conda} and~\eref{condb}) are satisfied, it does not mean that solution~\eref{2solution} (or solution~\eref{3solution}) has no singularity with the same conditions. In fact, one may observe the singular phenomena  in the near-field region $t\approx O(1)$ even if these conditions hold. Accordingly, Eqs.~\eref{cond1} and~\eref{cond2} (or Eqs.~\eref{conda} and~\eref{condb})) are just the \emph{necessary} conditions for solution~\eref{2solution} (or solution~\eref{3solution}) to be nonsingular.


\item[(ii)]

For the exponential multi-soliton solutions, the asymptotic solitons are usually localized in some straight lines where there exists a balance between two or more dominant exponential terms of the tau function~\cite{Biondini,XuJNMP}. But in deriving the mixed asymptotic solitons of solution~\eref{3solution}, we develop the asymptotic analysis method by considering the balance between some algebraic and exponential terms. It turns out that all the mixed asymptotic solitons of solution~\eref{3solution} are localized in some curves in the $xt$ plane. In comparison, there is a well agreement between the asymptotic expressions and solution~\eref{3solution} when $|t| \gg 1$. Therefore, such method is valid and may be applicable to studying the asymptotic behavior of multi-soliton solutions for other nonlocal evolution equations~\cite{Ablowitz2,vector,Sinha,DNLS,NKN,Fokas,NNWave,mKdV,SS,AB-KdV,Lou2}.

\end{enumerate}

\section*{Acknowledgement}
This work was supported by  the National Natural Science Foundation of China (Grant Nos. 11705284 and 61505054), by the Natural Science Foundation of Beijing Municipality (Grant No. 1162003), and by the Fundamental Research Funds of the Central Universities (Grant No. 2017MS051).

\end{document}